\documentclass[final,3p,times]{elsarticle}

\usepackage{graphicx}%
\usepackage{mathrsfs} 
\usepackage{multirow}%
\usepackage{amsmath,amssymb,amsfonts}%
\usepackage{amsthm}%
\usepackage{mathrsfs}%
\usepackage{float}
\usepackage[title]{appendix}%
\usepackage{xcolor}%
\usepackage{textcomp}%
\usepackage{manyfoot}%
\usepackage{booktabs}%
\usepackage{algorithm}%
\usepackage{algorithmicx}%
\usepackage{algpseudocode}%
\usepackage{listings}%
\usepackage{tikz}
\usepackage{graphicx}
\usetikzlibrary{shapes, arrows, positioning}
\usepackage{svg}
\usepackage{algorithm}
\usepackage{algpseudocode}
\usepackage{amsmath}
\usepackage{multicol}
\usepackage{array}
\usepackage{booktabs}
\usepackage{graphicx}
\usepackage{subcaption}
\usepackage{makecell}
\usepackage{booktabs} 
\usepackage{hyperref}
\usepackage{cleveref}

\newcommand{\imgpathchboo}{figs/}

\newcommand{\modifa}[1]{\textcolor{black}{#1}}
\newcommand{\modifb}[1]{\textcolor{black}{#1}}
\newcommand{\modifc}[1]{\textcolor{black}{#1}}
\newcommand{\degres}{^\circ}
\journal{Engineering Applications of Artificial Intelligence}

\begin{document}

\begin{frontmatter}

\title{Physics-Informed Neural Networks and Sequence Encoder: Application to heating and early cooling of thermo-stamping process}

\author[1]{Mouad ELAARABI} 
\author[1]{Domenico BORZACCHIELLO}
\author[2]{Philippe LE BOT}
\author[2]{Nathan LAUZERAL}
\author[1]{Sebastien COMAS-CARDONA}
\affiliation[1]{organization={Nantes Université, Ecole Centrale Nantes, CNRS, GeM, UMR 6183},
            city={Nantes},
            postcode={44321}, 
            country={France}}
\affiliation[2]{organization={Nantes Université, IRT Jules Verne},
                city={Bouguenais},
                postcode={44340}, 
                country={France}}

\begin{abstract}
In a previous work~\cite{elaarabi2025hybrid}, the Sequence Encoder for online dynamical system identification~\cite{SEQENCODER} and its combination with PINN (PINN-SE) were introduced and tested on both synthetic and real data case scenarios. The sequence encoder is able to effectively encode time series into feature vectors, which the PINN then uses to map to dynamical behavior, predicting system response under changes in parameters, ICs and BCs. Previously~\cite{elaarabi2025hybrid}, the tests on real data were limited to simple 1D problems and only 1D time series inputs of the Sequence Encoder. 
In this work, the possibility of applying PINN-SE to a more realistic case is investigated: \modifa{heating and early cooling} of the thermo-stamping process, which is a critical stage in the forming process of continuous fiber reinforced composite materials with thermoplastic polymer. The possibility of extending the PINN-SE inputs to multimodal data, such as sequences \modifb{of temporal 2D images} and to scenarios involving variable geometries, is also explored. The results show that combining multiple encoders with the previously proposed method~\cite{elaarabi2025hybrid} is feasible, we also show that training the model on synthetic data generated based on experimental data can help the model to generalize well for real experimental data, unseen during the training phase.
\end{abstract}


\end{frontmatter}
\section*{List of Abbreviations}

\begin{multicols}{2}
\begin{description}
  \item[AE] Absolute Error
  \item[APE] Absolute Percentage Error
  \item[BCs] Boundary Conditions
  \item[CNNs] Convolutional Neural Networks
  \item[FEM] Finite Element Method
  \item[GNNs] Graph Neural Networks
  \item[ICs] Initial Conditions
  \item[IoT] Internet of Things 
  \item[IR] Infrared
  \item[MAE] Mean Absolute Error
  \item[MAPE] Mean Absolute Percentage Error
  \item[MLPs] Multi-Layer Perceptrons
  \item[ODE] Ordinary Differential Equation
  \item[PA66] Polyamide 6-6
  \item[PA66GF] Polyamide 6-6 with Glass Fibers
  \item[PDE] Partial Differential Equation
  \item[PIML] Physics-Informed Machine Learning
  \item[PINN] Physics-Informed Neural Network
  \item[PINN-SE] Physics-Informed Neural Network with Sequence Encoder
  \item[PP] Polypropylene
  \item[PPGF] Polypropylene with Glass Fibers
  \item[QC] Quality of Contact
  \item[QPs] Query Points
  \item[RNNs] Recurrent Neural Networks
  \item[SE] Sequence Encoder
  \item[TCR] Thermal Contact Resistance
  \item[VAE] Variational Autoencoder
\end{description}
\end{multicols}

\section{Introduction}
The development of industrial processes by the integration of new technologies such as IoT, cloud computing, robotics and machine learning has led to significant improvements in automation, monitoring and control. \modifa{It enhanced} the efficiency of process by reducing defects and downtimes, while also improving overall quality. Real data plays a key role in any industrial process~\cite{pr10020335}. Different types of sensors can be employed to provide real-time visibility of process health~\citep{javaid2021significance}, such as temperature, pressure, proximity sensors and even visual data (images). These sensors are capable of capturing information ranging from scalar values (1D) to complex 3D spatial data. Each of these data types needs to be analyzed differently, while also considering their interconnections, in order to provide a full and clear picture of process performance. Handling diverse data streams from multiple sources in real time has become much easier with the development of IoT and its integration with cloud computing~\citep{breivold2017internet}. Such data can be used online via specific models to forecast process behavior and adjust parameters accordingly, or offline to gain a deeper understanding of process behavior with respect to parameters and external factors, which can then be exploited to improve the process itself.

The recent development of PIML~\citep{PIML, hao2023physics} has shown that including \modifb{prior physical  knowledge} of the process can improve model training and enhance generalizability. Different deep learning architectures have been developed to handle real-world data. For example, CNNs are widely used in diverse industrial contexts for image and video processing~\cite{schmidt2019deep}, RNNs for sequence data encoding and GNNs~\citep{scarselli2008graph} for graph-based data (which can also be applied to image processing, since an image can be represented as a structured graph). The incorporation of prior knowledge into these models can be achieved either by modifying their architecture or by adjusting the training objective. For instance, CNNs~\citep{maduranga2023symmetry} can be trained to preserve symmetry by adding a symmetry kernel to the architecture. NeuralODEs~\cite{chen2018neural} construct neural network architectures that follow the Euler discretization scheme of the target ODE. Alternatively, prior knowledge can be included directly in the training loss, as in the case of PINNs~\citep{PINN}. In this context, it is important to have models that can both integrate prior knowledge and also handle diverse data sources. As an example, in the study of flow over an espresso cup~\citep{cai2021flow}, PINNs were trained to encode sequences of 3D temperature fields and then use governing equations (Navier-Stokes and heat transfer) to predict the 3D velocity and pressure fields in continuous time. Similarly, CNNs have been combined with PINNs~\citep{hanna2024self} to extract local permeability during central injection molding, using images that capture the flow front (the propagation of liquid in the material). Other work~\cite{niaki2021physics} proposed an approach to tackle 3D thermomechanical problems using adaptive training of PINNs, while~\cite{qiu2023adaptive} studied the thermochemical evolution of \modifb{polymerization} in a composite material (thermoset polymer matrix) in contact with a tool. In the latter application, transfer learning was explored to minimize retraining time of PINNs when process parameters change. These studies highlight the potential of PINNs to leverage both data and physical constraints to achieve reliable predictions of thermo-stamping behavior. To sum up, incorporating image data into the training process provides a highly informative source. For this reason, the aim is to validate the model on its ability to encode such data.

\subsection{Problem definition}
The thermo-stamping process of composite~\cite{Chenreview2021} consists of heating a composite material plate above the thermoplastic matrix melting temperature, after which it is transferred into a mould where it is formed into the desired shape held and cooled to obtain a consolidated part with the required final geometry. During the process, different sources of information can provide feedback on process health, such as pressure or force sensors during the stamping phase and temperature sensors during the heating, stamping and cooling phases. Thermal analysis is a particularly important source of information, as temperature strongly influences the process. It can affect the friction mechanisms~\citep{brooks2022review, gong2020comprehensive} (ply-to-ply, ply-to-mold, inter-ply shears), as well as fiber movement (through the viscosity $\eta$ of \modifa{the polymer matrix}, which itself depends on sheet temperature), which may lead to defects during stamping. To this extent, temperature may be considered the most important processing parameter~\citep{brooks2022review}. 

\begin{figure}[!htbp]
    \centering
    \includegraphics[width=1\textwidth]{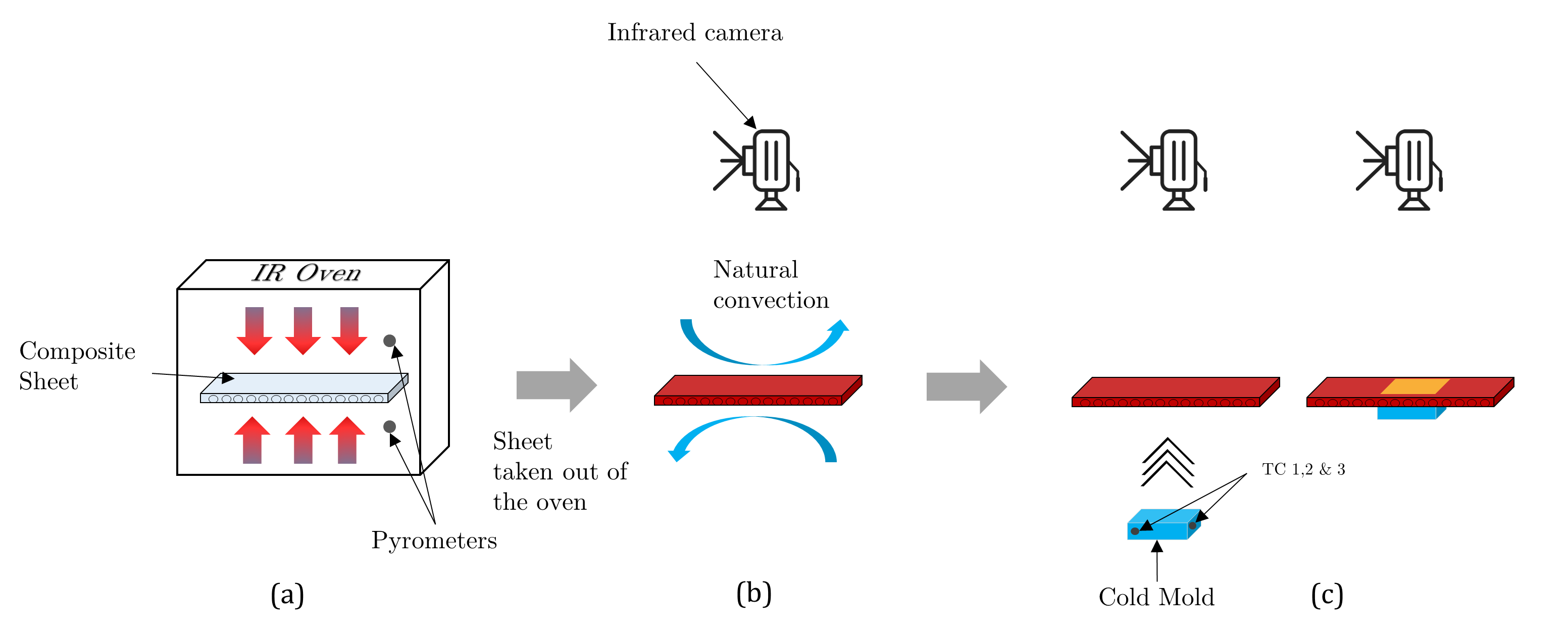}
    \caption{Overview of the setup used to validate PINN-SE. This setup includes the three main steps of the thermo-stamping process, focusing only on the thermal evolution. \modifc{An IR camera is mounted in a fixed position above the mold using a support arm, while the mold is actuated using a hydraulic cylinder and controlled by a control unit.} The steps can be summarized as follows: (a) The \modifa{sheet} is first heated to a certain temperature $T_h$ using an infrared (IR) oven. (b) The \modifa{sheet} is then taken out of the oven, \modifb{where it is exposed to natural convection during the transfer phase}. (c) Finally a cold mold is \modifa{raised to establish} thermal contact, emulating the cooling step of the stamping phase.}\label{process_overview_symbol}
\end{figure}

In this work, the aim is to validate the proposed method~\citep{elaarabi2025hybrid} in a scenario close to real world thermo-stamping, where data sources are diverse (time series, sequences of 2D images, geometries ...). The main objective is to verify whether PINN-SE can map changes in process parameters to dynamics behavior. The focus will be on the thermal part of the process, namely the heating, transfer and contact with the colder mold, without considering \modifa{the further forming stage}. This choice is made for simplicity and also because temperature is the main source of information in the process and has a major impact on product quality.  The main goal is to show that encoding different sources of information with PINN-SE is feasible, while accounting for variations in ICs, BCs and geometries and that it can handle spatio temporal problems using real data as input. The setup used in this study is presented in Fig.~\ref{process_overview_symbol} and can be summarized as follows:

\paragraph{Heating} Using an infrared oven, the composite sheet is heated to a surface target temperature $T_h$. Once this temperature is reached, the heating stops automatically.

\paragraph{Transfer} 
After heating, the sheet is taken out of the oven and exposed only to natural convection for 2 to 3 seconds. During this step, an IR camera is used to capture \modifa{the top surface} temperature of the sheet at $z=H$.  

\paragraph{\modifa{Mold \& sheet contact}} A mold (Fig.~\ref{process_overview_symbol}) is used to emulate the forming part of the thermo-stamping process. In this study, however, the focus is only on the cooling part of the sheet (the heat exchange between the low temperature mold and the hot sheet). During this step, thermocouples are used to measure the temperature of the mold at selected positions, while an IR camera captures \modifa{the top surface temperature} 2D field of the sheet.

\subsection{Paper contributions}
In this study, the aim is to use PINN-SE with CNNs to encode thermal images, thermocouple data and the positions of contact of the  Mold \& sheet (which may change between tests) during the convection part \modifa{(Fig.~\ref{process_overview_symbol}~(b))}, in order to predict the transient temperature during the contact phase \modifa{(Fig.~\ref{process_overview_symbol}~(c))}. The main contributions of this work are the following:
\begin{itemize}
    \item Extension of the application of PINN-SE to sequences of 2D thermal image real data
    \item Extension of PINN-SE to multi-encoder inputs
    \item Application of PINN-SE using real data from heating and early cooling of thermo-stamping
\end{itemize}
\subsection{Outline of the Paper}
To elaborate this work, this paper is divided as follows: in Sec.~\ref{Exp_part}, the experimental setup and procedure are presented, along with the samples used. Next in Sec.~\ref{Simu_part}, the synthetic data generation for PINN-SE training is presented. The model training, architecture and results are given in Sec.~\ref{Model_part}. The results of this work are then analyzed and discussed in Sec.~\ref{results_discc}. A conclusion and some perspectives of the work is then given in Sec.~\ref{Conc_part}.

\section{Approach for Generating Experimental Real Data}\label{Exp_part}
To validate the model, diverse conditions and materials are used. This means changing the heating temperature of the sheet, the boundary conditions by adjusting the initial temperature of the mold and the area/position of contact between the mold and the sheet. For the materials, two types of sheets with different polymers and different thicknesses are used. In this section, the samples, the details of the experimental process and the processing of the experimental data are presented.

\subsection{Materials}
\subsubsection{Samples}
For this experiment, two types \modifa{of sheets (Tab.~\ref{Summ_mat_data})} are used. A digital caliper was used to measure the thickness of each sample at its four corners and the thickness of each sample is defined as the average of these four measurements.

\begin{table}[!htbp]
    \caption{Summary of sheets: fiber type, polymer matrix, thickness, fabric type, warp-to-weft ratio and number of plies.}
    \label{Summ_mat_data}
    \centering
    \renewcommand{\arraystretch}{1.2} 
    \begin{tabular}{l l l c l c c}
        \hline
        \textbf{Material} & \textbf{Fiber} & \makecell{\textbf{Polymer}\\ \textbf{Matrix}} & \textbf{Thickness (mm)} & \textbf{Fabric} & \makecell{\textbf{Warp to } \\ \textbf{Weft Ratio}} & \makecell{\textbf{Number} \\ \textbf{of Plies}} \\
        \hline
        PA66GF & E-glass & PA66 & $1.5 \pm 0.008$ & Twill & 2/2 & 3 \\
        PPGF   & E-glass & PP   & $2.3 \pm 0.05$  & Twill & 2/2 & 2 \\
        \hline
    \end{tabular}
\end{table}

The properties of the \modifa{constituents} of each sheet are given in Tab.~\ref{tab:material_properties_chpf}. Both sheet types have nominal \modifa{square geometries of $25 \text{cm}\times25\text{cm}$}. In this experiment, 6 samples of \modifa{PA66GF sheets} and 4 samples of \modifa{PPGF sheets} are used. These samples were selected without any pre-requirements from a set of sheets. The main idea is to create a model based on a small number of samples.

\begin{table}[!htbp]
\caption{\modifa{Material properties of E-glass Fiber~\citep{berthereau2008fibres}, PP~\citep{mark2007physical, biron1998proprietes} and PA66~\citep{mark2007physical, biron1998proprietes}.}}
    \label{tab:material_properties_chpf}
    \centering
    \renewcommand{\arraystretch}{1.2} 
    
    \begin{tabular}{l c c c c}
        \hline
        \textbf{Property} & \textbf{E-glass} & \textbf{PP} & \textbf{PA66} & \textbf{Units} \\
        \hline
        Density ($\rho$) & 2580 & 910 & 1140 & kg/m$^3$ \\
        Specific Heat Capacity ($C_p$) & 830 & 1610 & 1460 & J/(kg$\cdot$K) \\
        Thermal Conductivity ($\lambda$) & 1.0 & 0.22 & 0.2 & W/(m$\cdot$K) \\
        \hline
    \end{tabular}
    
\end{table}

\subsubsection{Infrared Oven}
For the samples heating, an infrared oven is used, which can heat both sides (top and bottom) simultaneously \modifa{(SOPARA FOUR IRM 48 kW - 400 V)}. \modifc{This oven operates in the mid-infrared, around 2.6 micrometers~\cite{SOPARA2023}}, a technology that heats the surface of the sheet, which is then transferred inside the sheet by conduction. Two optical pyrometers are installed inside the oven, one on the top and one on the bottom. These sensors are used to capture local temperatures during heating.  An automated system is used to control the heating setup. It allows setting the target temperature $T_{\text{target}}$, the overall power level ($0$ - $100\%$) and controlling the top and bottom heating independently with different power percentages. The response time $t_{\text{resp}}$ (the time to reach the defined power percentages) and the holding time $t_{\text{hold}}$ (used to maintain the sheet at $T_{\text{target}}$) can also be configured. For all experiments, the response time $t_{\text{resp}}$ is set to it minimum value (1 s) and heating is stopped once $T_{\text{target}}$ is reached on both the top or bottom of the sheet.

\subsubsection{Infrared Camera}
For these tests, the ImageIR® 5300 from InfraTec GmbH is used. This camera can capture thermal images at a resolution of 320$\times$256, with a total of 81,920 detectors. It can be calibrated to measure temperatures over different ranges, with a theoretical accuracy of $99\% \pm 1\%$. The acquisition frame rate is set to 2 fps is used for all tests. The camera comes with IRBIS 3 software, which allows configuring internal parameters such as the emissivity of the target, the distance between the target and the camera and the ambient temperature to correct the measured temperatures. The software also enables capturing and saving thermographic images along with additional metadata, including the configured emissivity, ambient temperature and timespan for each snapshot.

\subsubsection{Mold}
\begin{figure}[!htbp]
    \centering
    \begin{subfigure}{0.45\textwidth}
        \includegraphics[width=0.7\textwidth]{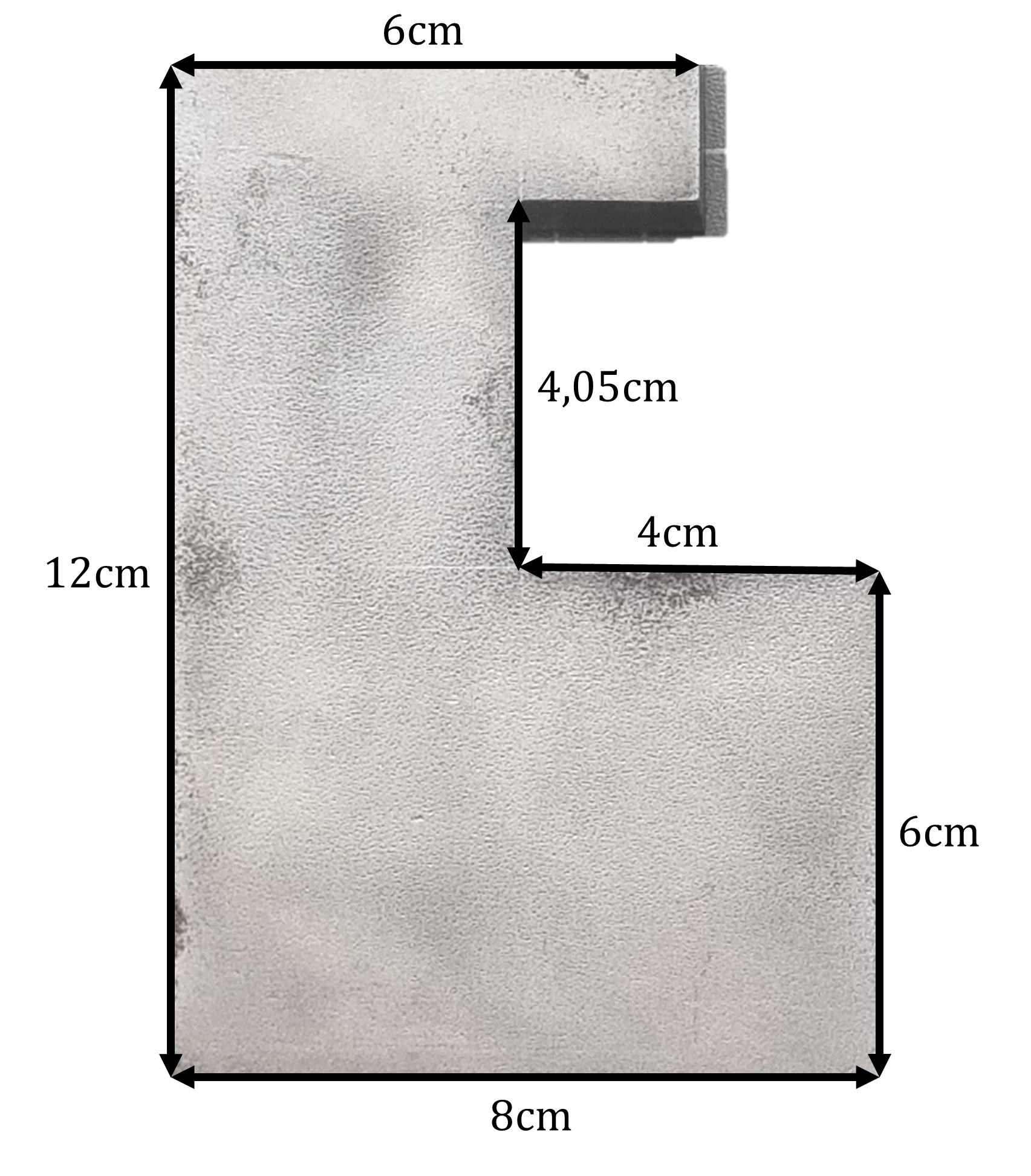}
        \caption{Mold geometry.}
        \label{fig:moldsub1}
    \end{subfigure}
    \hspace{0.05\textwidth} 
    \begin{subfigure}{0.4\textwidth}
        \includegraphics[width=0.6\textwidth]{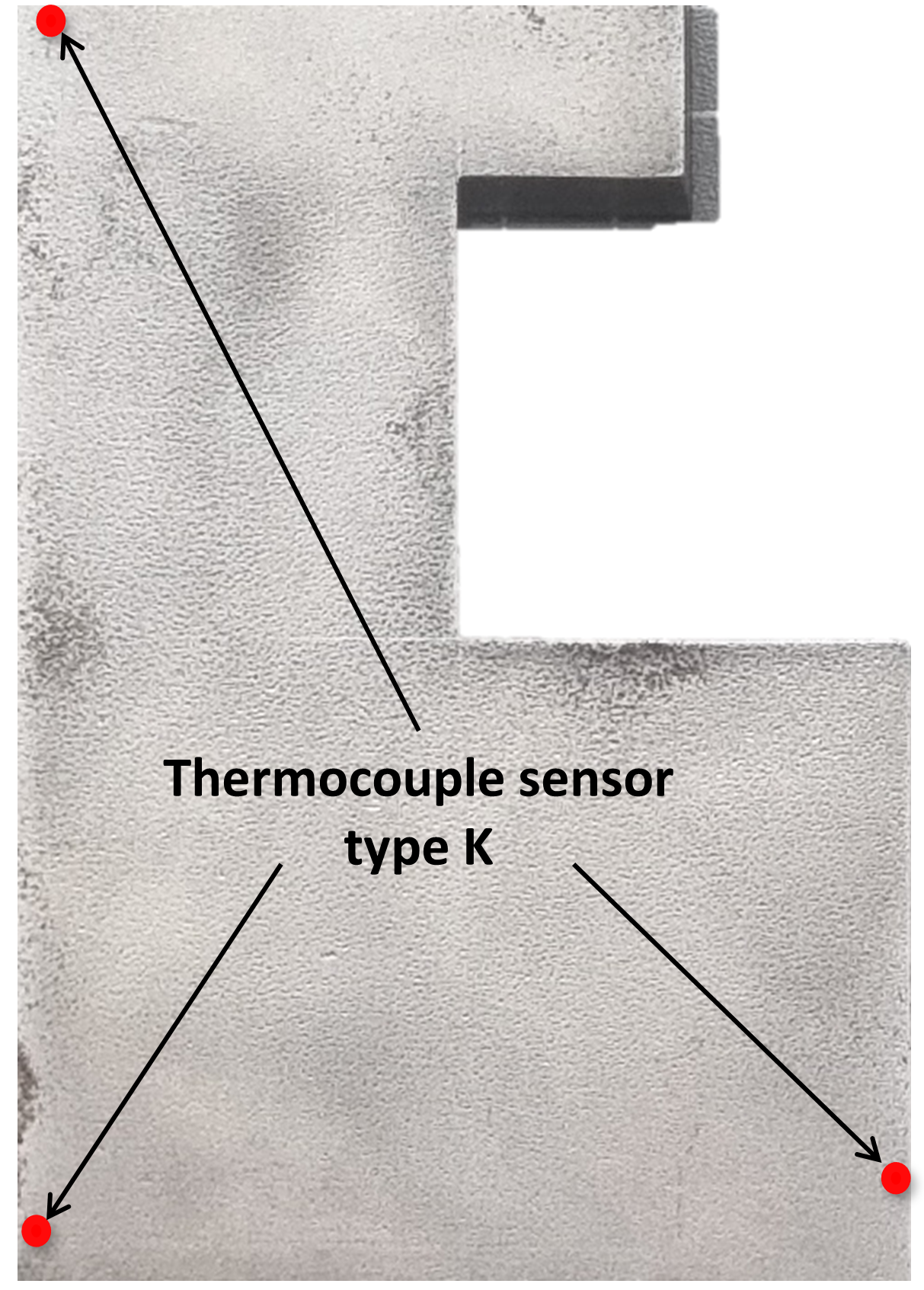} 
        \caption{The thermocouple sensors were positioned on the lateral surface of the mold and fixed in using polyimide heat resistant tape.}
        \label{fig:moldsub2}
    \end{subfigure}
    \caption{Top view of the mold geometry and thermocouple positions.}
    \label{fig.mold}
\end{figure}

A specific mold geometry (Top view in Fig.~\ref{fig:moldsub1}) with a thickness of 1.2~cm is used. The main idea was to employ a non-symmetrical shape in order to create edge effects during the thermal contact with the sheets as it occurs industrially. The mold is made of steel and three thermocouples were positioned close to edges in its lateral surface (Fig.~\ref{fig:moldsub2}) to measure the mold temperature before and during the contact with the sheets. 
The thermocouples are connected to a thermocouple box and acquired through the TracerDAQ software.

\subsection{Experimental setup}
\modifa{The experimental setup (mold, IR oven, IR camera) is displayed in Fig.~\ref{process_overview_symbol}. Data is recorded along the last two phases (b) and (c).}
The camera is mounted using a camera arm, which both fixes its position for all experiments and ensures that the camera view is perpendicular to the object of interest (sheets). The mold is mounted on a controlled hydraulic cylinder.
During the heating, the hydraulic cylinder is set to its \modifa{low position}. The sheet heating is controlled using the oven automation system: 100\% heating power is used and the oven is set to heat equally from top and bottom. The target temperature of \modifa{PA66GF} and \modifa{PPGF} sheets is set to $190~^\circ$C and $130~^\circ$ respectively.
Once the target temperature is reached, the heating stops and recording is started for both the camera and the mold thermocouples. The oven drawer is then slid out and during this phase there is no contact between the sheet and the mold. The sheet is exposed only to natural convection. A few seconds later (2-3 seconds), the mold is raised using the hydraulic cylinder, to establish contact with the sheet.
\begin{figure}[H]
    \centering
    \begin{subfigure}{\textwidth}
        \centering
        \includegraphics[width=0.8\textwidth]{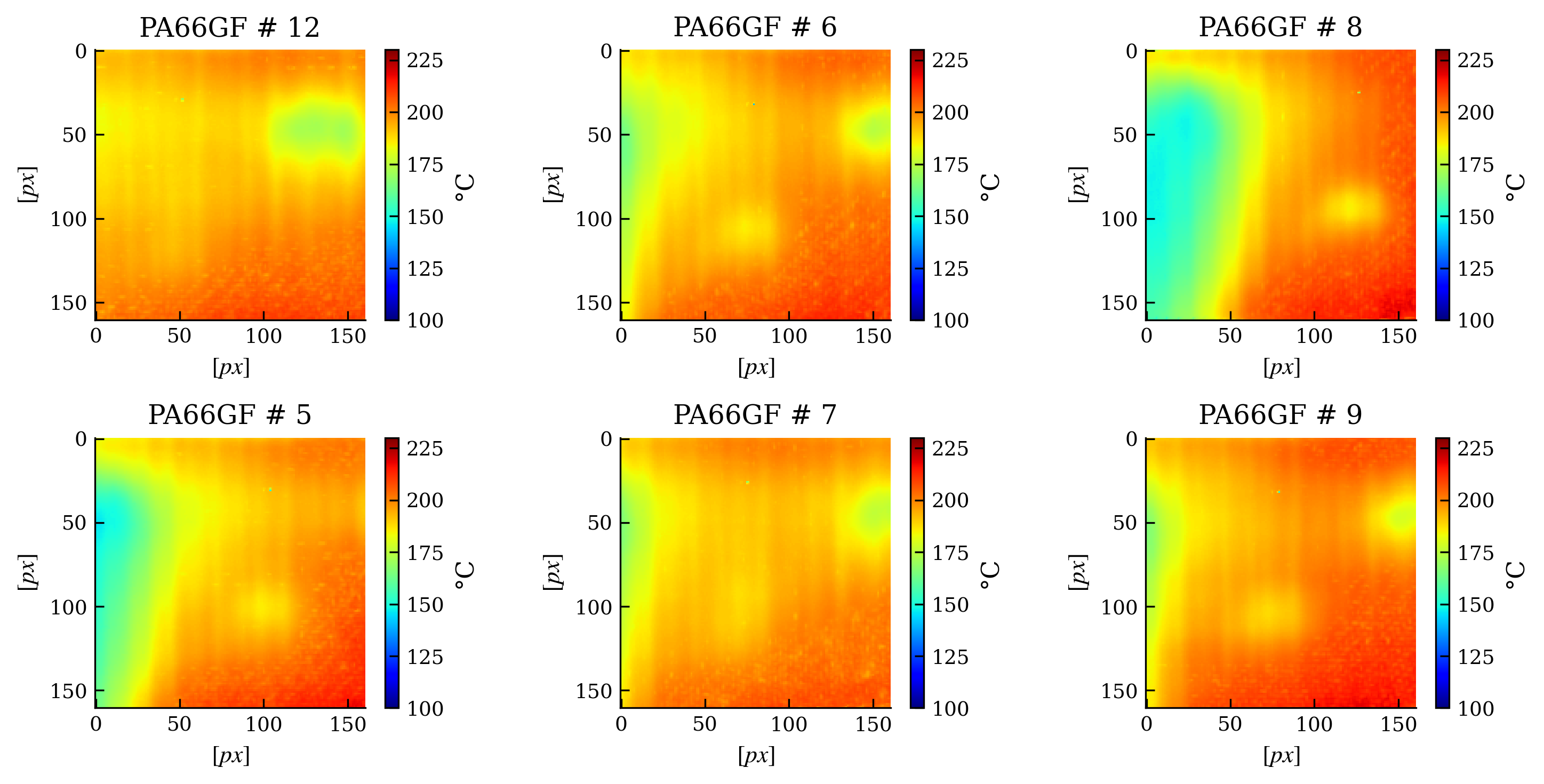}
        \caption{The temperature patterns at the surface after heating (phase (b) in Fig.~\ref{process_overview_symbol}) of the \modifa{PA66GF sheets}.}
        \label{fig:PA66_0s_contact}
    \end{subfigure}

    \vspace{0.5em}

    \begin{subfigure}{\textwidth}
        \centering
        \includegraphics[width=0.7\textwidth]{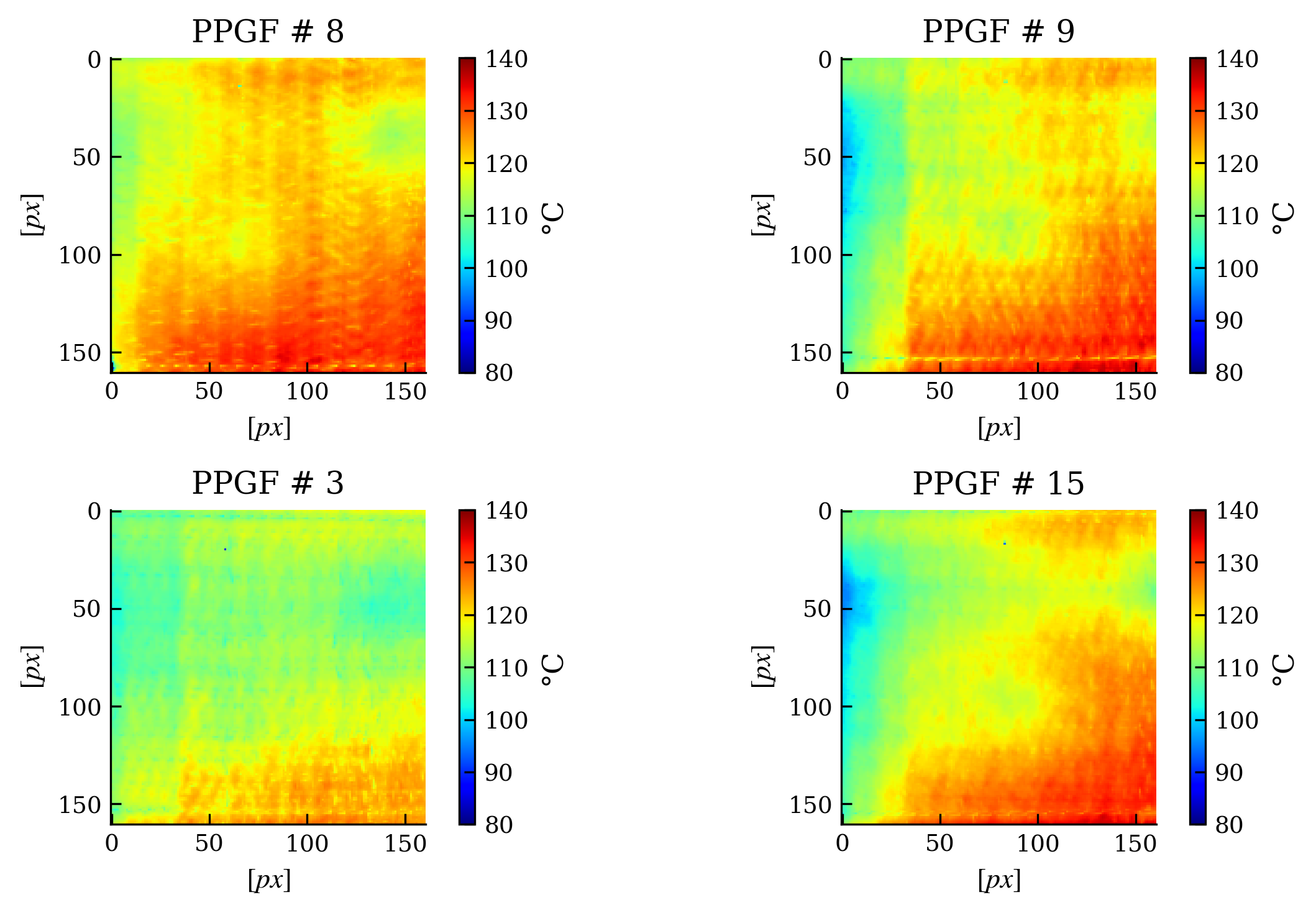}
        \caption{The temperature patterns at the surface after heating (phase (b) in Fig.~\ref{process_overview_symbol}) of the \modifa{PPGF sheets}.}
        \label{fig:PP_0s_contact}
    \end{subfigure}

    \caption{\modifa{Top view of Temperature patterns after heating of the PA66GF and PPGF sheets of the samples used in these experiments}.}
    \label{PA66_PP_0s}
\end{figure}

\subsection{Synthesis experimental data overview}\label{data_exp_synth}
\begin{figure}[H]
    \centering
    \begin{subfigure}{\textwidth}
        \centering
        \includegraphics[width=0.8\textwidth]{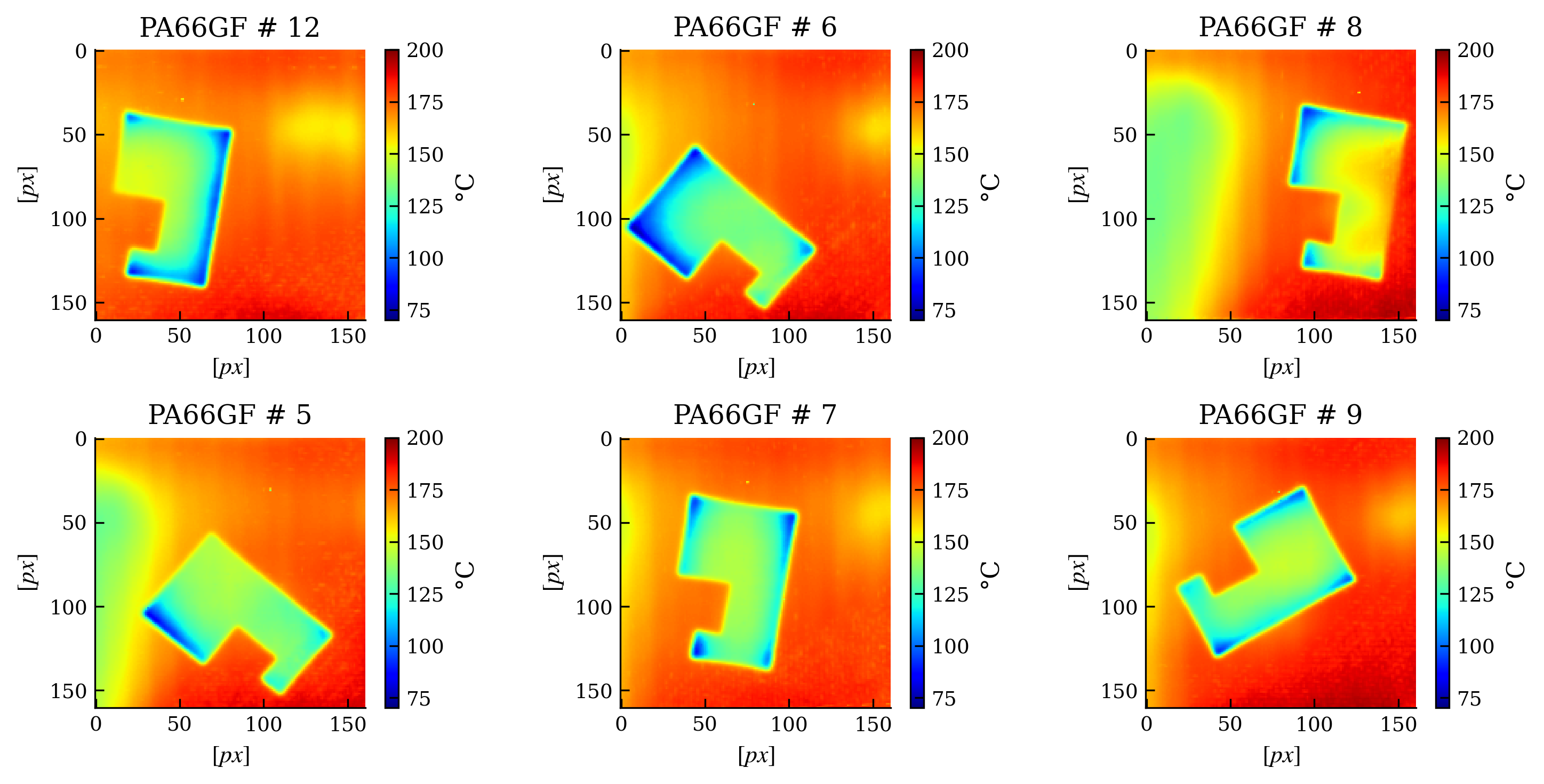}
        \caption{The temperature patterns at the surface 15 seconds after contact with the mold (phase (c) in Fig.~\ref{process_overview_symbol}) of the \modifa{PA66GF} sheets.}
        \label{fig:PA66_15s_contact}
    \end{subfigure}

    \vspace{0.5em} 

    \begin{subfigure}{\textwidth}
        \centering
        \includegraphics[width=0.7\textwidth]{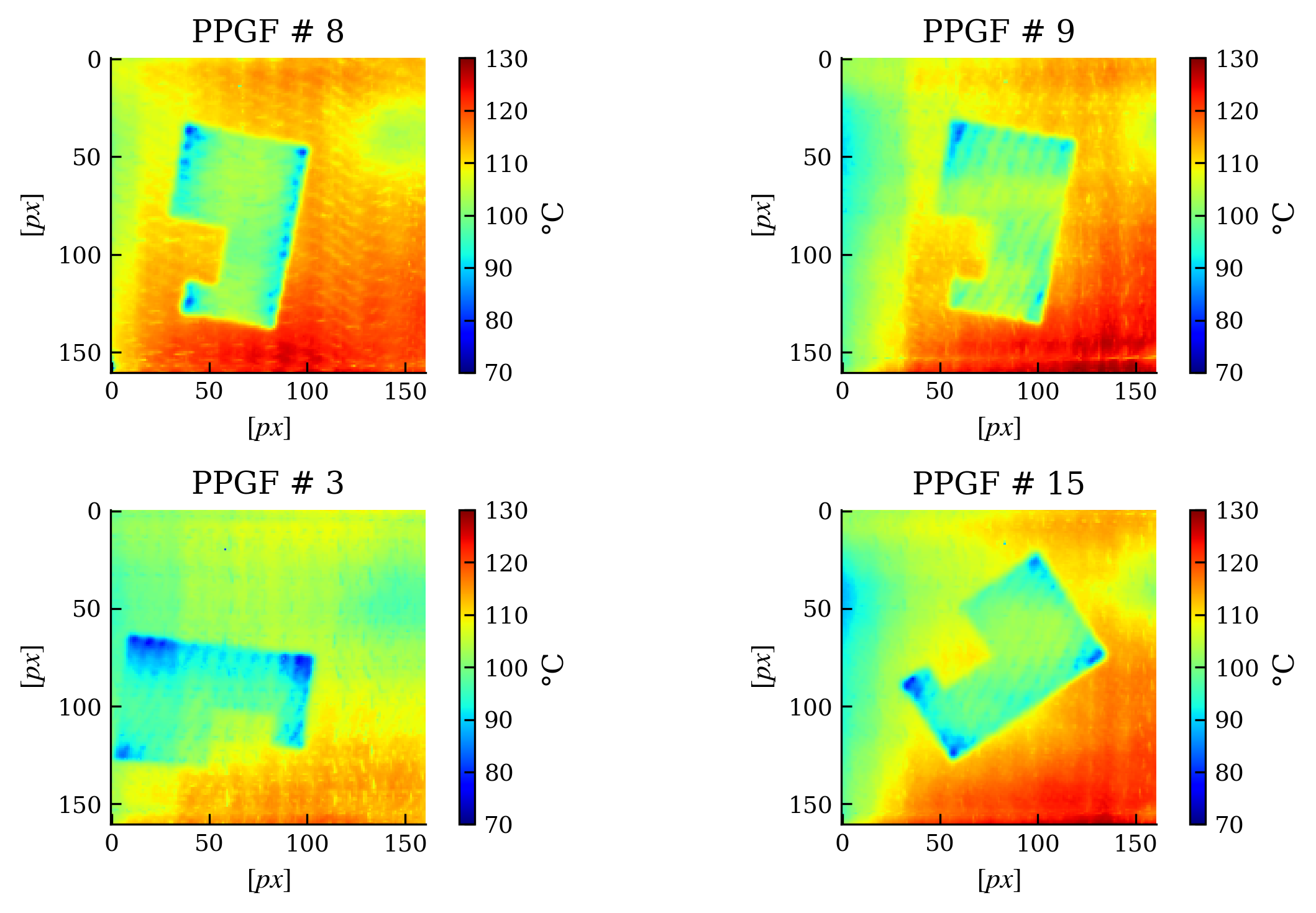}
        \caption{The temperature patterns at the surface 15 seconds after contact with the mold (phase (c) in Fig.~\ref{process_overview_symbol}) of the \modifa{PPGF} sheets.}
        \label{fig:PP_15s_contact}
    \end{subfigure}

    \caption{\modifa{Top view of temperature patterns at the surface 15 seconds after contact with the mold of the PA66GF and PPGF sheets for all the samples used in this experiments}.}
    \label{PA66_PP_15s}
\end{figure}

In this section, the process of data extraction and processing for each experiment is briefly introduced. The thermal images are logged as text files (2D floating matrices of $320\times256$) along with metadata such as the snapshot time, sequence order and the configuration parameters such as the set ambient temperature and emissivity. The thermocouple data are stored as  \textit{.csv} files with four columns, time and temperature at the three positions (Fig.~\ref{fig:moldsub2}).  
The main goal of this experiment is to generate a dataset under varied conditions, including the use of two different material sheets with distinct characteristics such as materials properties and thickness. \modifa{The top surface temperature} patterns are varied by translating horizontally the sheets in the IR oven, generating different thermal patterns (Fig.~\ref{PA66_PP_0s}) and by changing the contact position between the mold and the sheet (Fig.~\ref{PA66_PP_15s}).

\begin{figure}[!htp]
    \centering
    \includegraphics[width=0.9\textwidth]{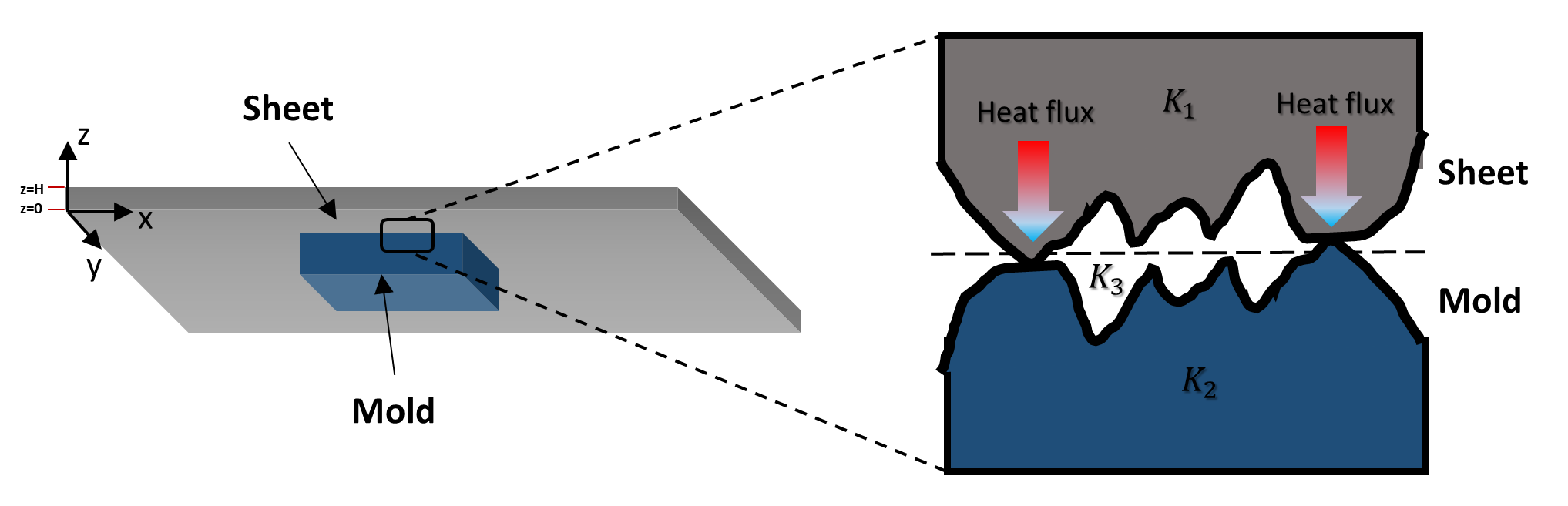}
    \caption{Illustration of heat transfer between two surfaces with asperities. The thermal contact resistance defines the quality of the contact and consequently the efficiency of heat transfer.}\label{overview_thermal_contact}
\end{figure}

During these experiments, the thermal contact resistance (TCR) (Fig.~\ref{overview_thermal_contact}), are expressed~\citep{delaunay2000nature} as :
\begin{equation}\label{TCR_Eq_def}
    R_s = \frac{T_\text{mold} - T_\text{sheet}}{\phi}, \quad
    [R_s] = \mathrm{m^2 \cdot ^\circ C / W}
\end{equation}
Where $T_{mold}$ and $T_{sheet}$ are the temperatures near the contact surface and $\phi$ is the heat flux within the contact interface. The TCR between the sheet and mold can impede heat transfer and thus reduce the efficiency of thermal contact. In this case, it is not negligible and is spatially variable (up to $40~^\circ\mathrm{C}$ difference observed across the sheet surface) and highly variable from one experiment to another, which adds a challenge to model validation. For polymer interfaces, this parameter depends on the surface roughness, the applied pressure, the gap thickness between the mold and sheet, as well as on time and temperature~\citep{gibbins2006thermal}. Estimating this parameter is crucial to evaluate the true heat transfer and, consequently, the quality of the stamping process. The boundary condition can be simplified as in Eq.~\ref{therma_resis_eq} for contact in the $z$ direction:
\begin{equation}\label{therma_resis_eq}
    -k \frac{\partial T}{\partial z} = h_c (T - T_\text{mold}),\quad (x,y)\in \Gamma_C
\end{equation}

\begin{figure}[!htbp]
    \centering
    \includegraphics[width=0.7\textwidth]{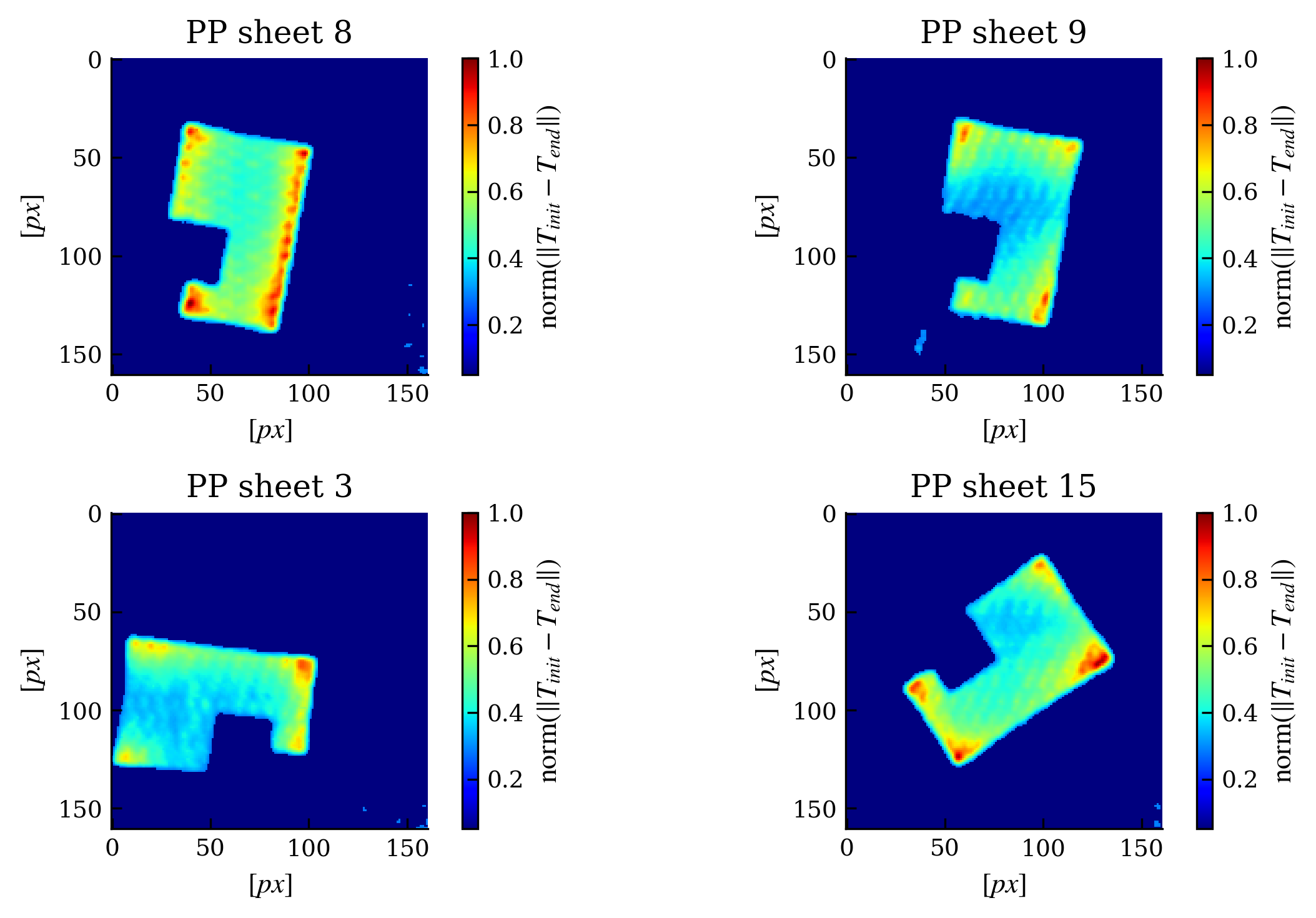}
    \caption{\modifa{Example of the top view normalized temperature difference}, $\text{normalized}(T^H_\text{init} - T^H_\text{end})$, for \modifa{PPGF sheets} used to approximate the contact heat transfer coefficient $h_c$.}\label{PP_thermal_quality_example}
\end{figure}

In this preliminary study, it is assumed that the TCR is constant (\modifa{in time and with respect to temperature}) and varies in position $(x,y)$. It is estimated using a numerical inverse method by simulating the contact between the mold and the sheet. The approach used defines the interface resistance as a function proportional to the temperature difference between two instances: 15 seconds after contact, $T^H_\text{end}$ and the initial temperature at the first contact with the mold, $T^H_\text{init}$, at the surface $z=H$. This temperature difference is then normalized using min-max scaling to have values between 0 and 1 (Fig.~\ref{PP_thermal_quality_example}). \modifa{In this study the heat transfer coefficient $h_c$ is approximated using this assumption as (Eq.~\ref{Thermal_res_prop})}:  
\begin{equation}\label{Thermal_res_prop}
    \begin{aligned}
        h_c &\propto (T^H_\text{init} - T^H_\text{end}) \\
        h_c &= 1/R_s\\
        R_s &= A~(1 - \text{normalized}(T^H_\text{init} - T^H_\text{end})) + B
    \end{aligned}
\end{equation}
\modifb{Where $R_s$ is the thermal contact resistance}. The choice of this function is based on the fact that the cooling of the sheet depends on the heat exchange between the mold and the sheet, which is highly influenced by $h_c$. For simplicity, a linear function is chosen to represent the proportional relationship between these variables.

\begin{figure}[!htp]
    \centering
    \includegraphics[width=1\textwidth]{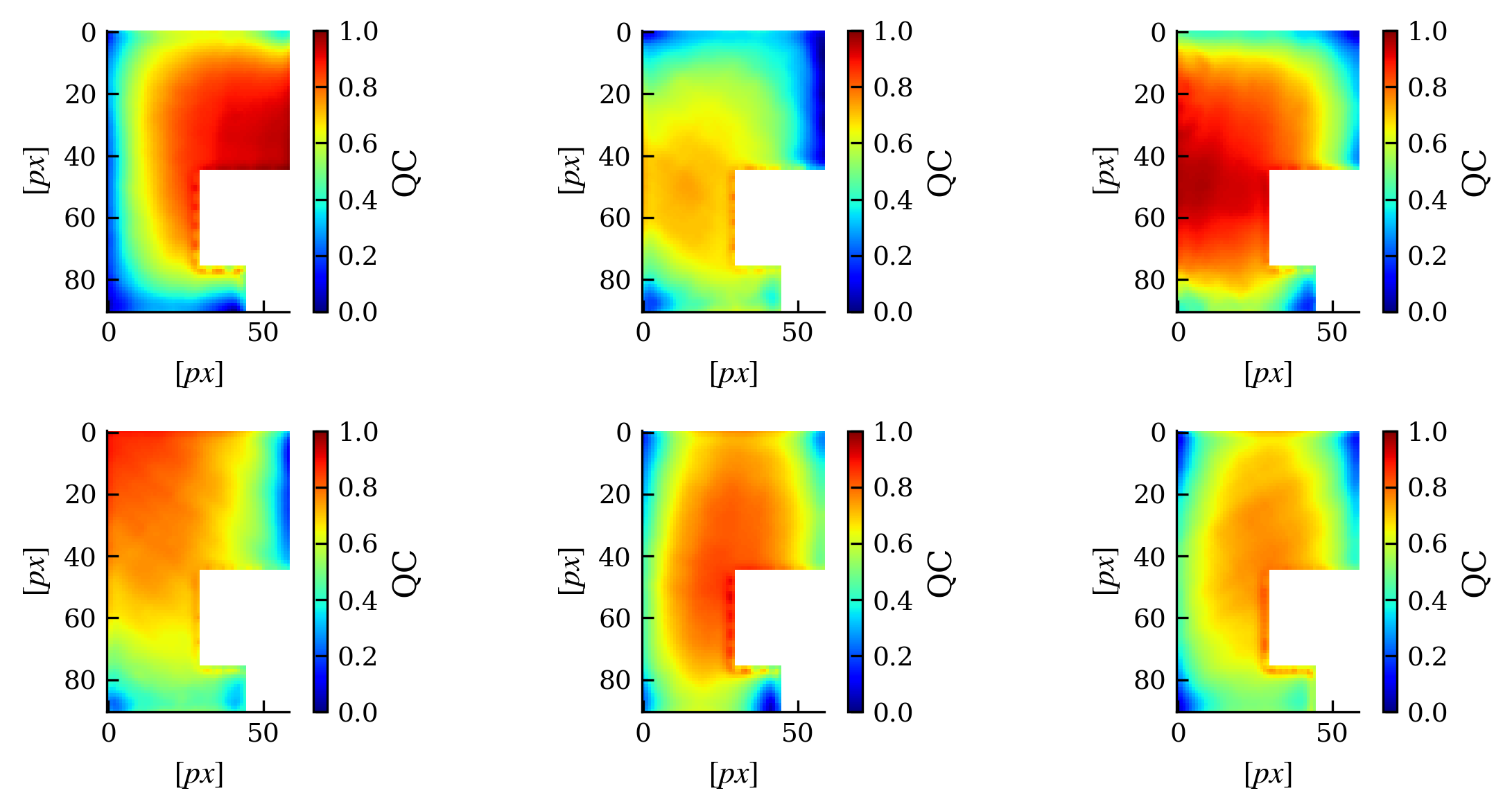}
    \caption{The Quality of Contact representation for all \modifa{PA66GF} sheets experiments.}\label{CQ_PA}
\end{figure}

\begin{figure}[!htp]
    \centering
    \includegraphics[width=0.85\textwidth]{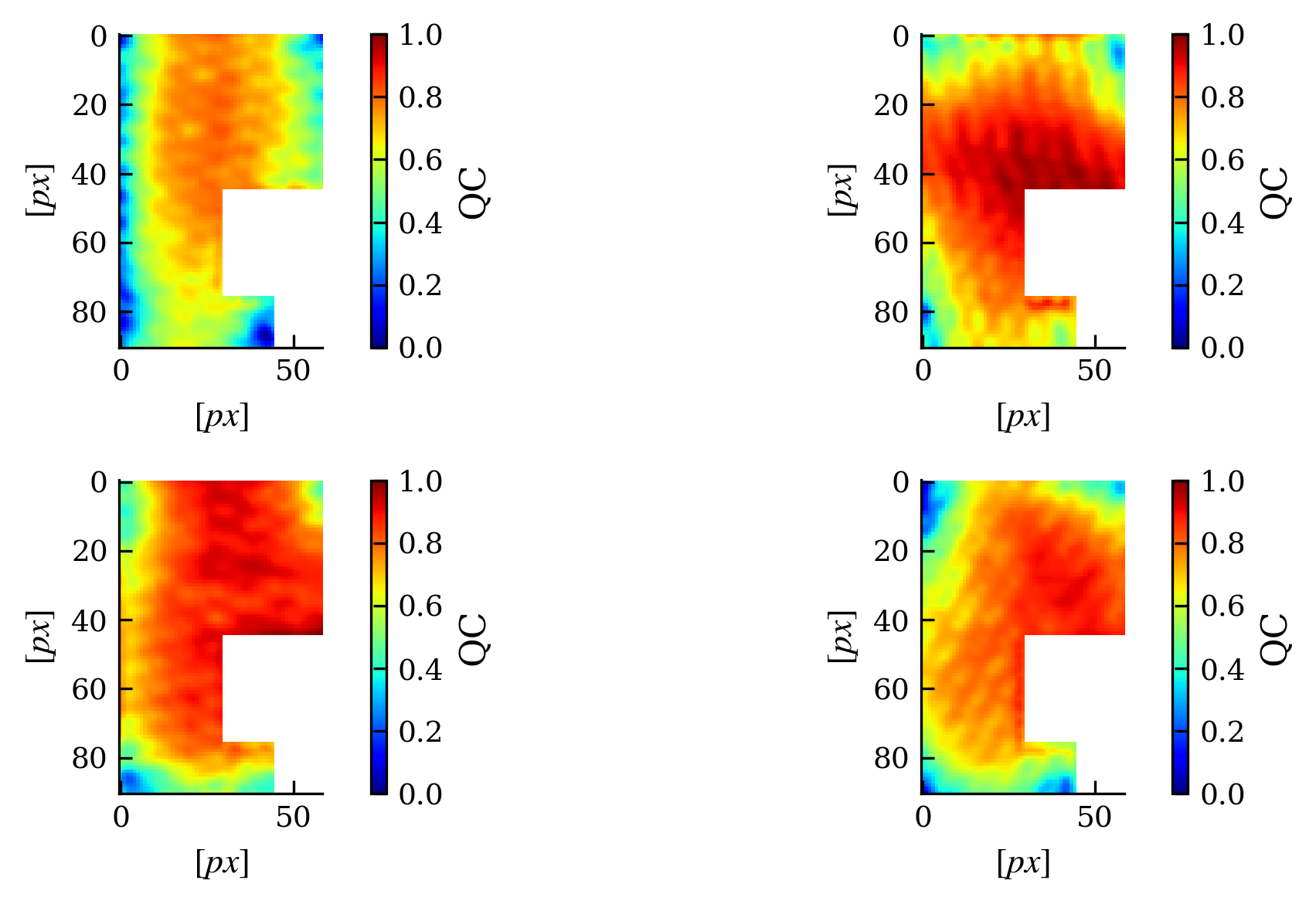}
    \caption{The Quality of Contact representation for all \modifa{PPGF} sheets experiments.}\label{CQ_PP}
\end{figure}

Using FEM in COMSOL 5.2 as the solver, the values of $A$ and $B$ (Eq.~\ref{Thermal_res_prop}) for \modifa{PPGF and PA66GF} sheets are identified separately by minimizing the average error (Eq.~\ref{inverse_propb_AB}) across all samples between the simulation and the surface temperature measured by the IR camera. 
\begin{equation}\label{inverse_propb_AB}
    \underset{A',~B'}{\text{argmin}} \quad \left|  T^H(t=15s,~\mathbf{x}; A,~B)_{sim}^{i} -  T^H(t=15s,~\mathbf{x})_{exp}^{i} \right|_1
\end{equation}
where $sim$ and $exp$ represent respectively the simulation and experimental data.
One optimal set of parameters is identified for each material: $A_\text{PP} = 140~\mathrm{m^2 \cdot ^\circ C / W}$ and $B_\text{PP} = 10^{-4}~\mathrm{m^2 \cdot ^\circ C / W}$ for all \modifa{PPGF} samples and $A_\text{PA} = 145~\mathrm{m^2 \cdot ^\circ C / W}$ and $B_\text{PA} = 1.05\times10^{-4}~\mathrm{m^2 \cdot ^\circ C / W}$ for \modifa{PA66GF} samples. 
For the sake of simplicity and to have a uniform image shape for model training (CNN), a crop is applied to both the \modifa{PA66GF} and \modifa{PPGF} image data to achieve a consistent shape. The cropped image has a square dimension of 20.7~cm, which will be the focus for the rest of this study. A summary of the QC for PA66GF and PPGF experiments is given in Figs.~\ref{CQ_PA} and~\ref{CQ_PP} respectively.

\modifa{To conclude, experiment data sets for 6 PA66GF sheets and 4 PPGF sheets have been acquired and the corresponding approximated coefficient $R_s$ and QC have been identified for diffirent sheet and mold positions.}

With all this variability, training the PINN-SE directly using limited amount of real data would be challenging. In the next section, we explain how synthetic data with similar patterns to the experimental data is generated and used to train the PINN-SE, which is then validated on experimental data.

\section{Generating synthetic data}\label{Simu_part}
\subsection{Simulation methodology}\label{simdetails}
The FEM simulation of the process is performed using COMSOL 5.2 software. The simulation is divided into two steps corresponding to the second and third steps (b and c) of the process (Fig.~\ref{process_overview_symbol}). The heating step is not simulated. \modifa{The top surface temperature} (e.g., Fig.~\ref{PA66_PP_0s}) from synthetic results is used as the initial condition for the simulation of the second and third steps of the process (Fig.~\ref{process_overview_symbol}). 

\paragraph{General configuration:} The initial temperature of the sheets is defined using a  \textit{.csv} file containing \modifa{the top surface temperature} at $z=H$. Then, using an interpolation function inside COMSOL, the temperature within the sheets is defined using a linear function. In this step, the material properties and geometries, including the mold positions relative to the sheets, are also defined using a  \textit{.csv} file containing the polygon coordinates corresponding to the mold nodes. The sheet thickness is defined as a variable parameter that can be easily changed from one simulation to another. Additionally, the quality of contact is configured using a \textit{.csv} file containing values between 0 and 1 and the coordinates of each value.

\paragraph{Part 1 of the simulation:} This corresponds to the second part (b) in Fig.~\ref{overview_thermal_contact}. In this phase, the heated sheet is exposed only to convection. For simplicity, the duration of this phase is fixed and set to 2 seconds for all simulations, with a heat \modifc{transfer coefficient $h = 10~\mathrm{W/(m^2.K)}$} and $T_{\infty} = 300~\mathrm{K}$. The initial condition is defined as described in the previous paragraph.  The 3D transient heat conduction equation for the sheet during this phase is defined as (Eq.~\ref{Thermal_PDE_EQ_ph1}):
\begin{equation}\label{Thermal_PDE_EQ_ph1}
    \begin{aligned}
        \displaystyle \frac{\partial T}{\partial t} &= \alpha \nabla^2 T, & (x,y,z)\in \Omega, \; t>0,\\[1mm]
        -k \frac{\partial T}{\partial z} &= h (T - T_\infty), & (x,y,z)\in \Gamma_R,\\
        -k \frac{\partial T}{\partial z} &= 0, & (x,y,z)\in \Gamma_I,\\
        T(t=0,x,y,z) &= T_0(x,y,z), & (x,y,z)\in \Omega.
    \end{aligned}
\end{equation}
Where $\alpha$ is the diffusivity, $\Gamma_R$ are the surface \modifc{where $z\in\{0,~H\}$ }and $\Gamma_I$ represent the lateral surface of the sheet.

\begin{figure}[!htbp]
    \centering
    \includegraphics[width=1\textwidth]{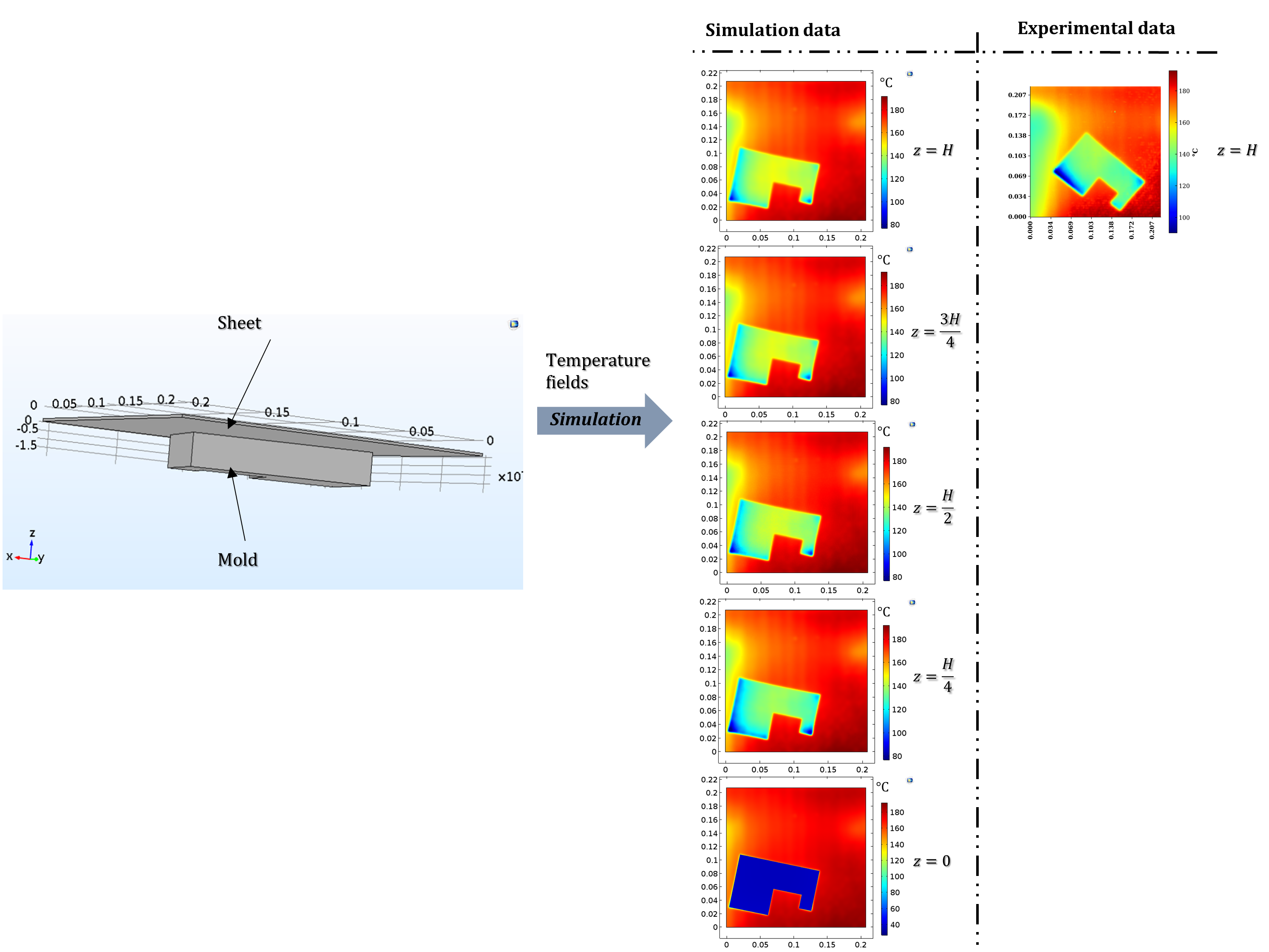}
    \caption{\modifa{Example of data generated by simulation versus a processed experimental data}.}\label{overview_data_gen}
\end{figure}

\paragraph{Part 2 of the simulation:} The results of simulation 1 (convection phase) are used as initial conditions to simulate the third part of the process, which corresponds to the contact between the sheet and the mold. In this phase, the initial temperature of the mold is a parameter that can vary between simulations. The duration of this phase is fixed to 10 seconds for all simulations.  The quality of contact is used to set $R_s$, which is read from a  \textit{.csv} file and varies from one simulation to another. For \modifa{PA66GF} sheet simulations, $A_{PA}$ and $B_{PA}$ are applied and similarly $A_{PP}$ and $B_{PP}$ for \modifa{PPGF} sheets.  The 3D transient heat conduction equation of the sheet during this phase is defined as (Eq.~\ref{Thermal_PDE_EQ_ph2}):
\begin{equation}\label{Thermal_PDE_EQ_ph2}
    \begin{aligned}
        \frac{\partial T}{\partial t} &= \alpha \nabla^2 T, & (x,y,z)\in \Omega, \; t>0,\\[1mm]
        -k \frac{\partial T}{\partial z} &= h (T - T_\infty), & (x,y,z)\in \Gamma_R, (z=H)\\
        -k \frac{\partial T}{\partial z} &= h_c (T - T_\text{mold}), & (x,y,z)\in \Gamma_C, (z=0)\\
        -k \frac{\partial T}{\partial z} &= 0, & (x,y,z)\in \Gamma_I,\\
        T(t=0,x,y,z) &= T_0(x,y,z), & (x,y,z)\in \Omega.
    \end{aligned}
\end{equation}
where $\Gamma_R$ represents the surfaces at $z=0$ (except the contact surface) and $z=H$ , $\Gamma_C$ represents the contact surface with the mold and $\Gamma_I$ represents the lateral surfaces of the sheet. \modifa{Some examples of simulation results are given in Fig.~\ref{overview_data_gen}, along with processed experimental data.}

\subsection{Data generations}\label{Data_gen_sim}
The number of experiments is very small for PINN-SE training and there is a lot of variability from one experiment to another. To overcome this limitation, data augmentation via simulation is used to generate a set of samples that matches the distribution observed in the real experiments. For each simulation the following were generated:

\paragraph{Surface temperature and QC patterns:} A combination of CNNs and Variational Autoencoders (VAE)~\citep{kingma2013auto} was used. This kind of hybrid (CNN+VAE) has been used in image generation tasks (face synthesis tasks,~\citep{hou2017deep}). In this work, a simple architecture was used to generate new surface temperature patterns (\modifa{Figs.~\ref{PA66_PP_0s}}) and new QC patterns (\modifa{Figs.~\ref{CQ_PA} and~\ref{CQ_PP}}). The idea is to train a probabilistic generative model that maps input data images $\mathbf{x}$ into a latent space $\mathbf{z}$ via an encoder $\mathrm{Encoder}(\mathbf{x})=q_\phi(\mathbf{z}\mid\mathbf{x})$ and reconstructs it using a decoder $\mathrm{Decoder}(\mathbf{z})=p_\theta(\mathbf{x}\mid\mathbf{z})$, such that $\mathbf{x}\approx\mathrm{Decoder}(\mathbf{z})$, where $\theta$ and $\phi$ are trainable parameters. Unlike deterministic autoencoders, the VAE is stochastic. The latent vector $\mathbf{z}$ is described by a distribution and the encoder outputs parameters of that distribution (commonly a mean and variance for each latent dimension). Training minimizes a reconstruction loss together with the Kullback-Leibler divergence between the approximate posterior $q_\phi(\mathbf{z}\mid\mathbf{x})$ and a chosen prior $p(\mathbf{z})$ (normal distribution $\mathcal{N}(\mathbf{0},\mathbf{I})$), which regularizes the latent space to be consistent with training distribution. The decoder then maps samples from the latent distribution back to data space, producing outputs that follow the training data distribution. Two CNN-VAE models are employed, one for temperature patterns and another for QC patterns. Both encoder and decoder networks are CNN-based. The temperature images shown in Figs.~\ref{PA66_PP_0s} are normalized separately for \modifa{PA66GF} and \modifa{PPGF} (each material normalized with respect to its own dataset) to scale values into $[0,1]$. After normalization, one VAE is trained for the temperature patterns (\modifa{based on both PA66GF and PPGF data}). The same procedure is applied to QC images.

\begin{figure}[!htbp]
    \centering
    \includegraphics[width=0.7\textwidth]{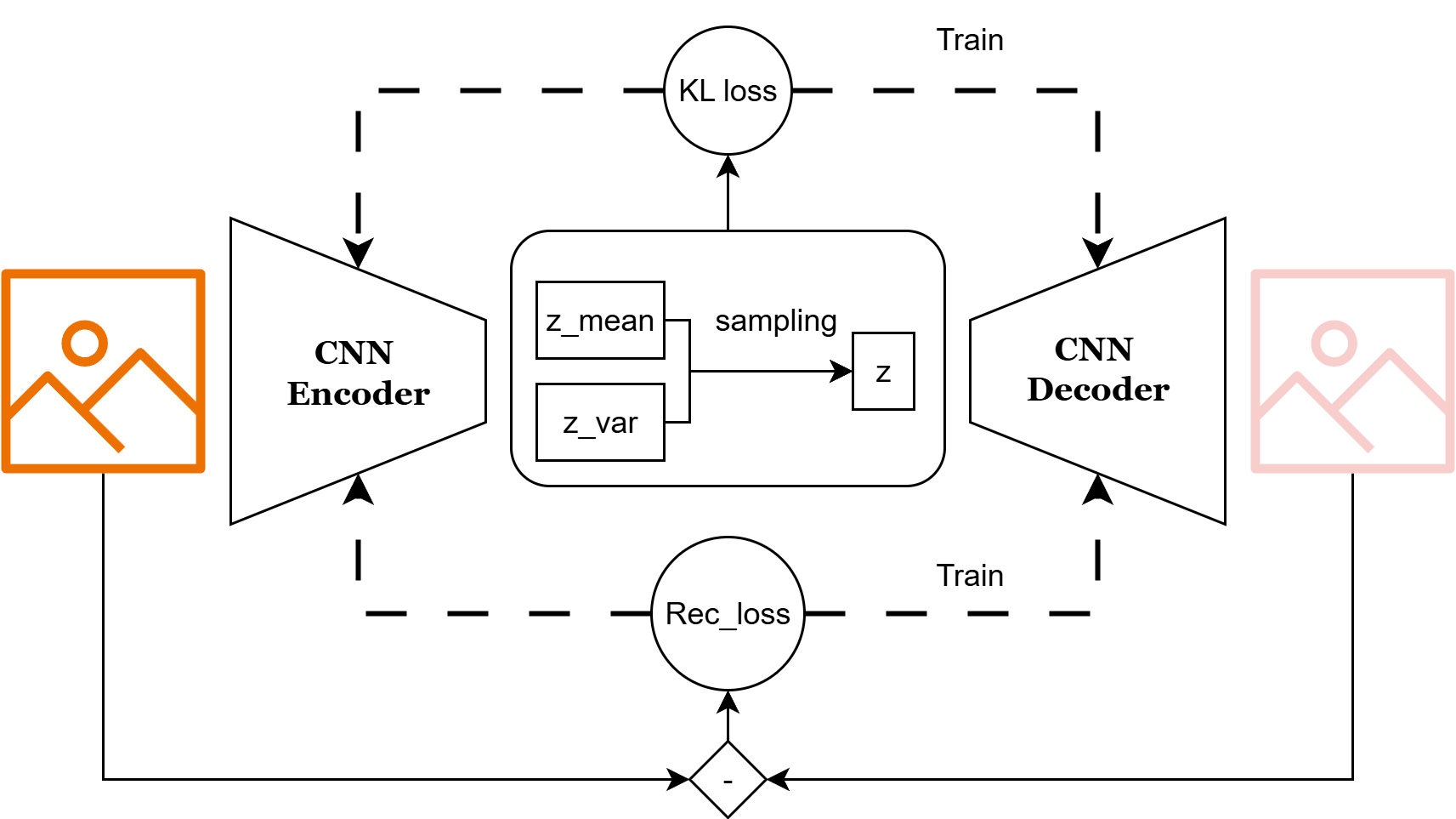}
    \caption{Overview of the CNN-VAE architecture.}\label{VAE_MODEL}
\end{figure}
\begin{figure}[!htp]
    \centering
    \includegraphics[width=0.85\textwidth]{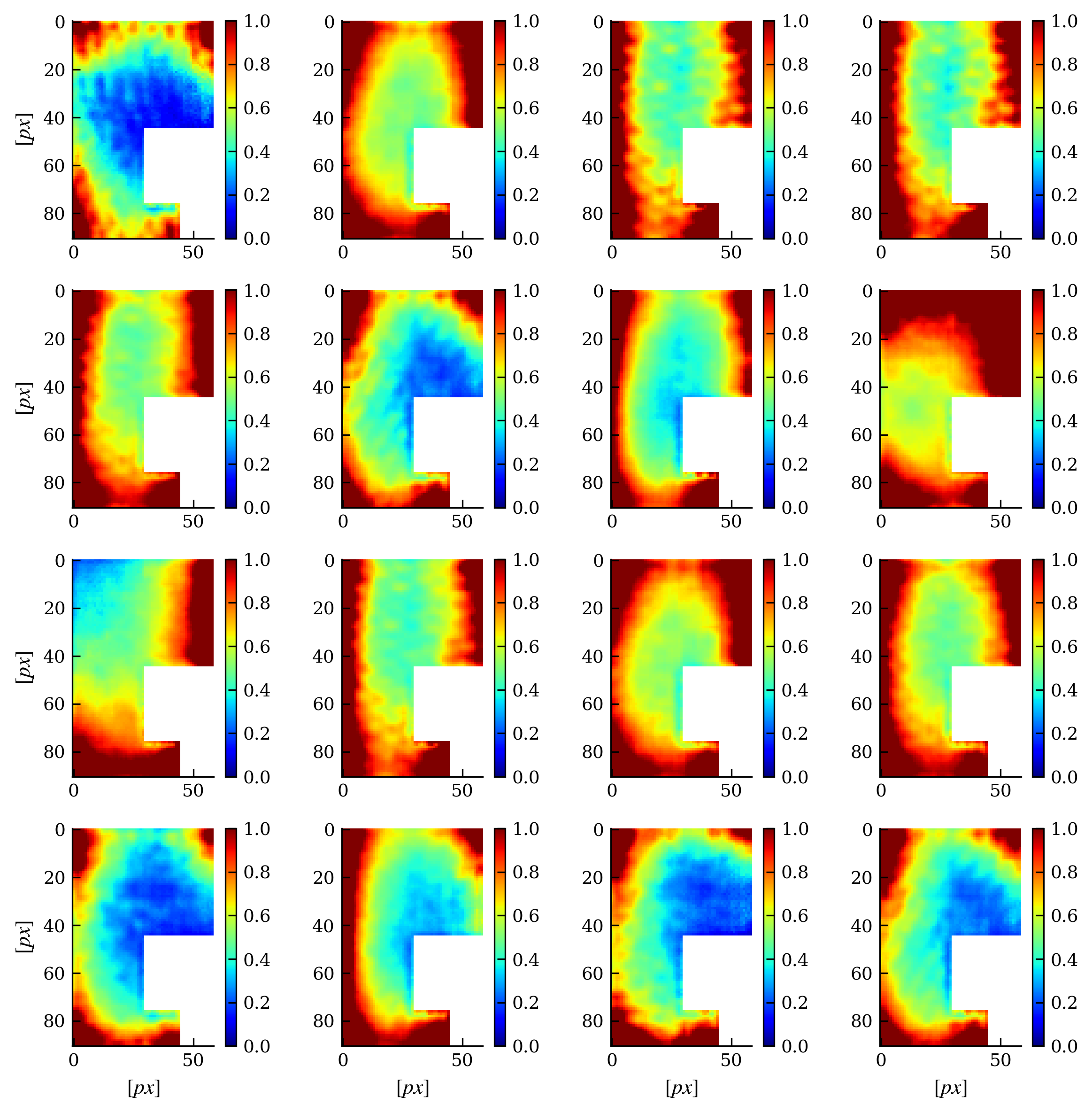}
    \caption{Examples of the QC generated using the VAE model.}\label{QC_VAE}
\end{figure}

Once the models (Fig.~\ref{VAE_MODEL}) are trained, latent vectors are sampled from a normal distribution and fed into the decoder to generate new datasets. Some examples of QC datasets produced by the trained VAE model are shown in Fig.~\ref{QC_VAE}. The details of the architecture are provided in~\ref{app_ch3_3}. The generated QC and temperature are then exported to \textit{.csv} files, which are subsequently used to run the FEM simulations as explained in Sec.~\ref{simdetails}.

\paragraph{Mold position with respect to the sheet:} The contact position between the sheet and the mold is generated randomly. A set of polygon coordinates (representing the mold with respect to the sheet) is created and these configurations are then used to both \modifa{PA66GF} and \modifa{PPGF} simulations.  
\paragraph{Material properties:} Different material properties are obtained by varying the fiber volume fraction $V_f$ of \modifa{PA66GF} and \modifa{PPGF} within $\pm 5\%$ since the nominal value of $V_f$ may slightly vary from one sheet to another.  
\paragraph{Initial temperature of the mold:} The initial mold temperature at the first instant of contact with the sheet is selected randomly between $25 \degres C$ and $35 \degres C$ for both \modifa{PA66GF} and \modifa{PPGF} simulations, which corresponds to the real experiments data.  

A combination of these variations is then used to generate 100 simulations for \modifa{PA66GF sheets} and 100 simulations for \modifa{PPGF sheets}, which are then used to train the model. For each simulation, the configuration used is saved, which includes \modifa{the top surface temperature}, the QC, the mold position, the sheet thickness and the thermal properties, alongside the simulation results. For each simulation, the results of the first part (Sec.~\ref{simdetails}) are stored at 5 instances, \modifc{defined as $t_{prior} = \tfrac{kt_0}{4} \quad \text{where} \quad k \in [0,1,\ldots,4]$, where $t_0=2 s$,} at $z=H$ with a surface grid of $161 \times 161$. For the second part of the simulation, the temperature is saved at 11 instances, \modifc{defined as $t_{post} = \tfrac{kt_f}{10} \quad \text{where} \quad k \in [0,1,\ldots,10]$, at 5 positions $z = \tfrac{kH}{4} \quad \text{where} \quad k \in [0,1,2,3,4]$, where $t_f=10 s$,} with a surface grid of $161 \times 161$. \modifa{An example of simulation generated data and experimental data is given in Fig.~\ref{overview_data_gen}}.  
This dataset is then used to train the PINN-SE model. In the next Sec.~\ref{Model_part}, the data preparation process for model training is detailed alongside the model architecture and its training strategy, with results on both the validation dataset and real data.

\section{Integrating CNNs, Deep Sets and PINNs for Thermal Monitoring}\label{Model_part}
The raw data generated above from the simulations require pre-processing before being used to train the PINN-SE, which is described in Sec.~\ref{data_prep_pinnse}. The model architecture is then introduced in Sec.~\ref{pinn_se_archi}, followed by a discussion of the training strategy in Sec.~\ref{traing_strategy} and finally, the results are presented and analysed in Sec.~\ref{results_discc}.

\subsection{Data Preparation}\label{data_prep_pinnse}
Each simulation is divided into input and output. The input data is further split into two types: the first corresponds to the encoder part of the model (details in Sec.~\ref{pinn_se_archi}), which in this case includes the simulation results of the convection phase only (first part, Sec.~\ref{simdetails}), the QC corresponding to the simulation, the initial temperature of the mold, the thickness of the sheet and the coordinates of the mold polygon. The second type of input corresponds to the PINN part, which is the space-time points at which the temperature is to be predicted. The output is the temperature at these space-time points. After structuring each simulation into input-output datasets, the data is divided into three sets, a training set with 70\% of the simulations (140 datasets) randomly selected and two 15\% sets for testing and validation, respectively.  
In this study, it is assumed that the quality of contact (QC) is known. The main objective is to validate whether the model can capture the changes in QC and update its predictions accordingly.

\subsection{Model Architecture}\label{pinn_se_archi}

\begin{figure}[!htp]
    \centering
    \includegraphics[width=0.9\textwidth]{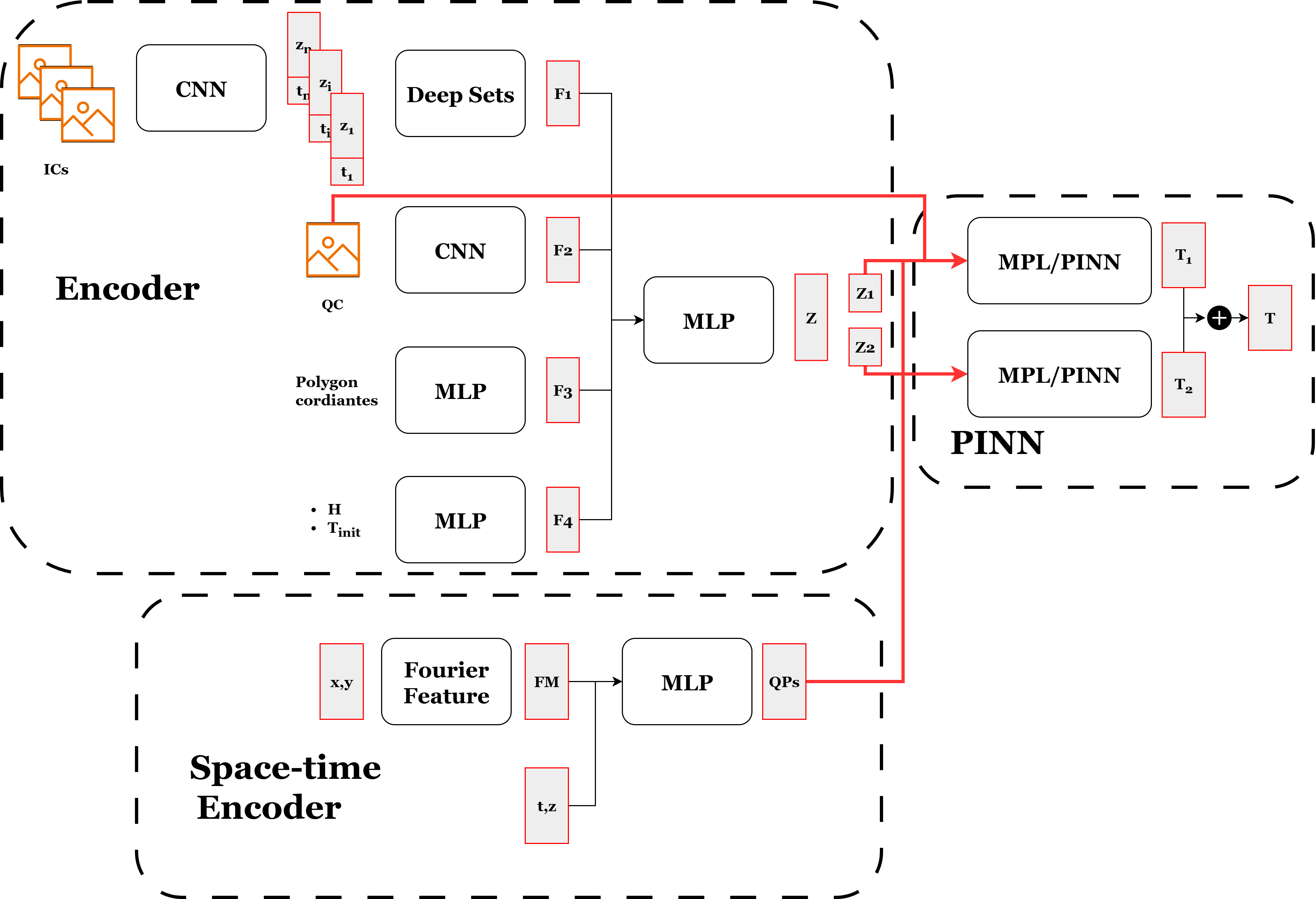}
    \caption{An overview of the PINN-SE architecture used in this application. The PINN-SE can be divided into three parts. The Encoder is composed of four sub modules: the first encodes a sequence of images representing the temperature during the convection phase of the process (part (b) in Fig.~\ref{process_overview_symbol}) using a CNN and Deep Sets. The second model is a CNN that encodes the Quality of Contact, $\text{QC}\in\mathbb{R}^{H\times W\times C}$. The third and fourth models are simple MLPs that encode mold coordinates, sheet thickness and initial contact temperature. Each model produces a feature vector $F_i\in\mathbb{R}^{B\times f_i}$, $i\in\{1,2,3,4\}$, which are concatenated to form $Z\in\mathbb{R}^{B\times(\sum_{i=1}^4 f_i)}$. The second main part is the Space-time Encoder, which encodes spatio-temporal data into Query Points using Fourier feature layers~\citep{tancik2020fourierfeaturesletnetworks} and an MLP. The final part is the PINN, which takes the concatenation of QPs and $Z$ as input. In this work, two MLP are used and their outputs are summed to give the final temperature prediction.
    }\label{PINN_SE_ARCHi}

\end{figure}

The model architecture overview is given in Fig.~\ref{PINN_SE_ARCHi}, while the detailed architecture is provided in~\ref{app_ch3_4}. The Encoder part of the model comprises four components. The first is a CNN that encodes \modifa{the top surface temperature} during the convection phase (first part Sec.~\ref{simdetails}, corresponding to part (b) in Fig.~\ref{process_overview_symbol}). This CNN encodes a sequence of surface temperature images representing the cooling of the sheet due to convection. Each image has dimensions $\text{ICs} \in \mathbb{R}^{H\times W}$, with $H = 161$ and $W = 161$. The CNN then generates a feature vector for each image in the sequence. This feature vector is concatenated with its corresponding time value, adding temporal information and enabling the Deep Set to capture time dependencies. A Deep Set is then used to encode the temporal dimension of these latent vectors. This approach was used in~\cite{chai2023driver} to encode CNN outputs from a sequence of driver images for head pose detection. The process can be summarized as in Eq.~\ref{eq_encode_cnndpst}:

\begin{equation}\label{eq_encode_cnndpst}
    \begin{aligned}
        \textbf{CNN}(T(t=t_i)) &= z_i\in\mathbb{R}^{B\times fv_1} \quad\text{where}\quad T_i\in\mathbb{R}^{B\times161\times161}\quad t_i\in\{0,~0.5,~1,~1.5,~2\}\\
        \text{concatenate}(\{t_i,z_i\}) &= \textbf{F}_{vect}\in\mathbb{R}^{B\times N_{t}\times (fv^{'}_1+1)} \quad\text{where}\quad N_{t}\ =\text{nbr of thermal images} \\
        \textbf{DeepSet}(\textbf{F}_{vect}) &= F_1\in\mathbb{R}^{B\times f_1}
    \end{aligned}
\end{equation}
Where $B$ is the batch size (the number of sets per training step). The second component of the encoder is a CNN that encodes the QC (with the coordinates of each pixel). The input dimension of QC fed to the model is $\text{QC} \in \mathbb{R}^{H\times W\times C}$, where $H = 91$ and $W = 59$ represent the height and width of the image and $C = 3$ represents the number of channels. The first channel corresponds to the QC values, while the second and third channels represent the corresponding spatial coordinates (Eq.~\ref{eq_encode_cnnqc}):
\begin{equation}\label{eq_encode_cnnqc}
    \textbf{CNN}(QC) = F_2\in\mathbb{R}^{B\times f_2}  
\end{equation}
An MLP is used to encode the polygon coordinates (8 points in total, each with $x$ and $y$ coordinates, 16 values in total). The last encoder is an MLP for the initial temperature of the mold and the thickness of the sheets. Each encoder generates a feature vector $F_i \in \mathbb{R}^{B \times f_i}$, with $i \in \{1,2,3,4\}$, which are then concatenated to produce $Z \in \mathbb{R}^{B \times \sum_{i=1}^4 f_i} = Z\in \mathbb{R}^{B \times f}$.
The second part of the model encodes the space-time input data before feeding it to the PINN. Fourier feature layers~\citep{tancik2020fourierfeaturesletnetworks} are used to encode the $x$ and $y$ dimensions. The output is then concatenated with $z$ (space) and time $t$ and fed into an MLP that encodes all spatio-temporal data to generate Query Points (QPs). Fourier feature layers are applied only to $x$ and $y$ because the temperature varies significantly in these dimensions, which helps the model converge faster. This step can be summarized as in Eq.~\ref{eq_encode_time_space}:
\begin{equation}\label{eq_encode_time_space}
    \begin{aligned}
        \text{FFl}(\textbf{x}) &= \text{FM}\in\mathbb{R}^{B\times N_{\textbf{x}}\times f_{fm}}, \quad \textbf{x}=(x,y)\\
        \text{concatenate}(\text{FM}, z,t) &= \text{QPs}_{prior} \in \mathbb{R}^{B\times N_{\textbf{x}}\times (f_{fm}+2)}\\
        \text{MLP}(\text{QPs}_{prior}) &= \text{QPs} \in\mathbb{R}^{B\times N_{\textbf{x}}\times f_{qps}}
\end{aligned}
\end{equation}

Where $\text{FFl}$ represents the Fourier feature layer and it encoding dimension is $f_{fm}$  and $N_{\mathbf{x}}$ the number of space-time points at which the temperature is predicted. In this work, two MLPs are used to represent the PINNs. The predictions of both models are summed to give the final temperature prediction. This approach was tested in~\cite{elaarabi2025hybrid} and it showed very good results for model convergence.  Each MLP takes as input the concatenation of the Query Points (QPs) and $Z_j \in \mathbb{R}^{B \times f'}$ where $j \in \{1,2\}$ and $2f' = f$. Both $Z_j$ have the same dimension and therefore $f$ needs to be even.  
An essential step that helps model convergence is the creation of a direct connection between the QC and the QPs. This is done by calculating the distance between the QPs and the QC coordinates, which then allows the model to approximate the QC at the QPs. This approach is inspired by attention mechanisms, where the model encodes the relationships between its inputs. With this approach, each QP has a different weight, represented by the QC values.

\subsection{Model Training Process}\label{traing_strategy}
For each simulation (element of batch) the encoder model takes as input the QC, thickness, initial temperature of the mold, the polygon coordinates and a sequence of thermal images of the convection phase. Since the Deep Set can handle different sequence lengths, the model is trained with different thermal image sequence lengths (1-5 thermal images), which allows predictions and updating as new data comes in. All encoder inputs are normalized to values between 0-1, which helps the model training convergence. For each batch, the time-space encoder is used to encode the ICs, BCs points and PDE collocation points. Note here that the PINN is used to solve only the thermal contact \modifa{step of the process} (Eq.~\ref{Thermal_PDE_EQ_ph2}), while the convection \modifa{step} data are used by the encoder (such as sequence of thermal images).  
To train the model, a similar approach as in~\cite{elaarabi2025hybrid} is used. The loss function to optimize is defined as (Eq.~\ref{loss_TH_4D}):
\begin{equation}\label{loss_TH_4D}
    \begin{aligned}
    \mathcal{L}_{\text{total}} &= \beta(\mathcal{L}_{\text{SUP}}  + \mathcal{L}_{\text{ICs}} +  \mathcal{L}_{\text{BCs}}) + \lambda \mathcal{L}_{\text{PDE}} + \gamma \mathcal{L}_{\text{l1}}, \\
    \mathcal{L}_{\text{SUP}} &= \|T(t, \textbf{x}, z=\frac{H}{2}) - \widetilde{T}(t, \textbf{x}, z=\frac{H}{2})\|_2 \quad \mathbf{x} = (x,y)\\
    \mathcal{L}_{\text{ICs}} &= \|T(t=0, \textbf{x}, z) - \widetilde{T}(t=0, \textbf{x}, z)\|_2 \\
    \mathcal{L}_{\text{BCs}} &= \|T(t, \textbf{x}, z) - \widetilde{T}(t, \textbf{x}, z)\|_2 \quad z \in {0, H} \\
    \mathcal{L}_{\text{PDE}} &= \|\frac{\partial T}{\partial t} - \alpha \nabla^2 T\|_1\\
    \end{aligned}
\end{equation}
Where $\mathcal{L}_{\text{SUP}}$ is the supervised loss. In this work, the temperature at $z=\frac{H}{2}$ is used to help model training. This is not mandatory but it helps the model convergence speed. $\mathcal{L}_{\text{PDE}}$ is the physics residual, $\mathcal{L}_{\text{ICs}}$ is the initial condition loss and $\mathcal{L}_{\text{BCs}}$ represents the boundary condition loss. For simplicity, for the training Dirichlet conditions are used instead of flux based conditions, since the simulations enforce the original flux conditions. The parameters $\beta$ and $\lambda$ are used to give different weights to the losses. In this work $\beta = 10$, while $\lambda$ is dynamically variable depending on the supervised loss $\mathcal{L}_{\text{SUP}}$. The term $\mathcal{L}_{\text{l1}}$ represents the L1 regularization applied to all layers of the model, with a regularization coefficient $\gamma = 10^{-6}$. The model training strategy is summarized in Algo.~\ref{alog_3D_HT}.

\begin{algorithm}
    \caption{Training strategy of PINN-SE for 3D transient heat conduction equation.}\label{alog_3D_HT}
    \begin{algorithmic}[1]
    \State \textbf{Initialize parameters:}
    \State \( \beta = 10 \) 
    \State \( \lambda = 0 \) 
    \State first\_time\_met = \text{True} 
    \State Initialize learning rate \( \text{lr} = 10^{-3} \)
    
    \For{each epoch \( e \) in training}
        \State Compute loss components: $\mathcal{L}_{\text{SUP}}, \mathcal{L}_{\text{ICs}}, \mathcal{L}_{\text{BCs}}, \mathcal{L}_{\text{PDE}}$, $\mathcal{L}_{\text{l1}}$
        \State Compute total loss: $\beta(\mathcal{L}_{\text{SUP}}  + \mathcal{L}_{\text{ICs}} +  \mathcal{L}_{\text{BCs}}) + \lambda \mathcal{L}_{\text{PDE}} + \gamma \mathcal{L}_{\text{l1}}$ 
        
        \State \textbf{Update physics loss parameter:}
        \If{Sup loss $<$ thr0}
            \If{first\_time\_met} 
                \State first\_time\_met = False
                \State $\lambda$ = $10^{-8}$
            \Else
                \State $\lambda = \lambda \times (1 + 10^{-3})$
                \State $\lambda = \min(1, \lambda)$
            \EndIf
        \Else
            \If{not first\_time\_met} 
                \State $\lambda = \lambda \times (1 - 10^{-2})$
                \State $\lambda = \max(10^{-8}, \lambda)$
            \EndIf
        \EndIf
        
        \State \textbf{Update learning rate:}
        \If{lr $>$ $2\times 10^{-4}$} 
            \State lr = lr $\times$ $9.9992 \times 10^{-1}$
        \ElsIf{lr $>$ $10^{-4}$} 
            \State lr = lr $\times$ $9.9999 \times 10^{-1}$
        \EndIf
        \If{\( e \mod 5000 = 0 \)}
            \State lr = lr $\times$ $9.0 \times 10^{-1}$
        \EndIf
    \EndFor
    
    \end{algorithmic}
\end{algorithm}

\subsection{Results and Discussion}\label{results_discc}
To train the model, 70\% of the dataset is selected randomly. An optimal model is selected based on the test loss results. For the model validation, 15\% of the dataset unseen during the training is kept. The model is validated based on its prediction at $z = \frac{kH}{4} \quad k\in\{0,1,2,3,4\}$ for synthetic data (validation dataset) and the model prediction at $z=H$ is also validated for real data. To assess the model prediction error, two metrics are used, the $\textbf{MAPE}$ (Mean Absolute Percentage Error) and $\textbf{MAE}$ (Mean Absolute Error). Since the data has no outliers and is not intermittent, without zero values, using $\textbf{MAPE}$ is a suitable metric to validate the model~\citep{FORCASTBP}, while the $\textbf{MAE}$ gives the real temperature error of the model.

\paragraph{Validation on synthetic data:} The model predictions are evaluated using thermal image sequences of varying lengths (2 to 5 frames) as input to the encoder (Fig.~\ref{PINN_SE_ARCHi}). The assessment is performed at five discrete positions along the thickness, $z = \frac{kH}{4}$ with $k \in \{0,1,2,3,4\}$ (\modifa{Fig.~\ref{overview_data_gen}}), over the entire surface $\mathbf{x}=(x,y)\in[0,0.207]\times[0,0.207]$ discretized on a $161\times161$ grid and at five time instances $t = \frac{k\modifc{t_f}}{10}$, $k\in\{0,2,5,7,10\}$. The $\textbf{MAPE}$ and $\textbf{MAE}$ are calculated for each dataset as in Eq.~\ref{APE_def_eq} and Eq.~\ref{AE_def_eq}:

\begin{equation}\label{APE_def_eq}
    \begin{aligned}
        \mathrm{APE}(\mathbf{x};~t_i,z_i) &= \frac{|\hat{T}(\mathbf{x},z_i,t_i)-T(\mathbf{x},z_i,t_i)|}{|T(\mathbf{x},z_i,t_i)|}\times 100,\\
        \mathrm{MAPE}_j &= \frac{1}{|\mathcal{X}|}\sum_{\mathbf{x}\in \mathcal{X}} \mathrm{APE}_j(\mathbf{x}), \quad d_j \in \mathcal{D}_{\mathrm{val}}, \quad j = 1, \dots, N_{\mathrm{val}}\\[1mm]
    \end{aligned}
\end{equation}

\begin{equation}\label{AE_def_eq}
    \begin{aligned}
        \mathrm{AE}(\mathbf{x};~t_i,z_i) &= |\hat{T}(\mathbf{x},z_i,t_i) - T(\mathbf{x},z_i,t_i)|, \\
        \mathrm{MAE}_j &= \frac{1}{|\mathcal{X}|}\sum_{\mathbf{x}\in \mathcal{X}} \mathrm{AE}_j(\mathbf{x}), \quad d_j \in \mathcal{D}_{\mathrm{val}}, \quad j = 1, \dots, N_{\mathrm{val}}\\[1mm]
    \end{aligned}
\end{equation}
Where $\hat{T}$ represents the model prediction and $T$ presents the simulation results (ground truth). To provide a comprehensive picture of model accuracy, the mean, the 50\% (median) and 90\% percentiles of $\textbf{MAPE}$ are reported over all validation datasets. The same procedure is applied to the $\textbf{MAE}$ metric. The definitions of these metrics are given in Eq.~\ref{MAPE_def_eq} for $\textbf{MAPE}$ and Eq.~\ref{MAE_def_eq} for $\textbf{MAE}$.

\begin{equation}\label{MAPE_def_eq}
    \begin{aligned}
        Q_q(\mathrm{MAPE}(t_i,z_i)) &= \inf \Big\{ \eta : \frac{1}{|\mathcal{D}_{\mathrm{val}}|}\sum_{d_j \in \mathcal{D}_{\mathrm{val}}}\mathbb{I}(\mathrm{MAPE}_j \le \eta) \ge q;\quad d_j \in \mathcal{D}_{\mathrm{val}}, \quad j = 1, \dots, N_{\mathrm{val}} \Big\}, \; q\in\{0.5,0.9\},\\
        \mathrm{MAPE}(t_i,z_i) &=\frac{1}{N_{\mathrm{val}}}\sum_{d_j\in \mathcal{D}_{\mathrm{val}}} \mathrm{MAPE}_{j}.\\
    \end{aligned}
\end{equation}

\begin{equation}\label{MAE_def_eq}
    \begin{aligned}
        Q_q(\mathrm{MAE}(t_i,z_i))& =\inf \Big\{ \eta : \frac{1}{|\mathcal{D}_{\mathrm{val}}|}\sum_{d_j \in \mathcal{D}_{\mathrm{val}}}\mathbb{I}(\mathrm{MAE}_j \le \eta) \ge q;\quad d_j \in \mathcal{D}_{\mathrm{val}}, \quad j = 1, \dots, N_{\mathrm{val}} \Big\}, \; q\in\{0.5,0.9\},\\[1mm]
        \mathrm{MAE}(t_i,z_i) &= \frac{1}{N_{\mathrm{val}}}\sum_{d_j\in \mathcal{D}_{\mathrm{val}}} \mathrm{MAE}_j.\\
\end{aligned}
\end{equation}

\begin{figure}[!htbp]
    \centering
    \includegraphics[width=0.85\textwidth]{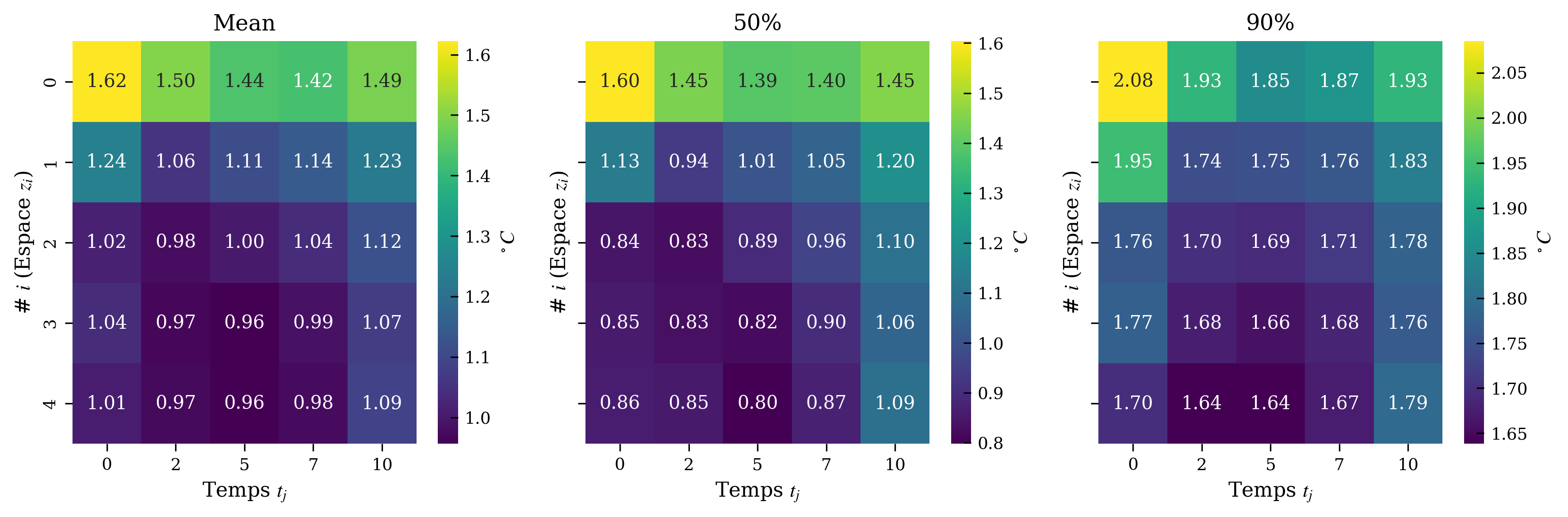}
    \caption{\modifa{Model performance across different spatial locations ($z = \frac{kH}{4} \quad k\in\{0,1,2,3,4\}$ and $t=\frac{k\modifc{t_f}}{10} \quad k\in \{0,2,5,7,10\}$), using \textbf{MAE} metrics and using only 2 thermal images as input to the encoder. The figure summarizes the mean, median and 90th percentile errors of the model, as defined in Eq.~\ref{MAE_def_eq}, across all validation datasets.}}\label{results_AE_chp3}
\end{figure}

\begin{figure}[!htbp]
    \centering
    \includegraphics[width=0.85\textwidth]{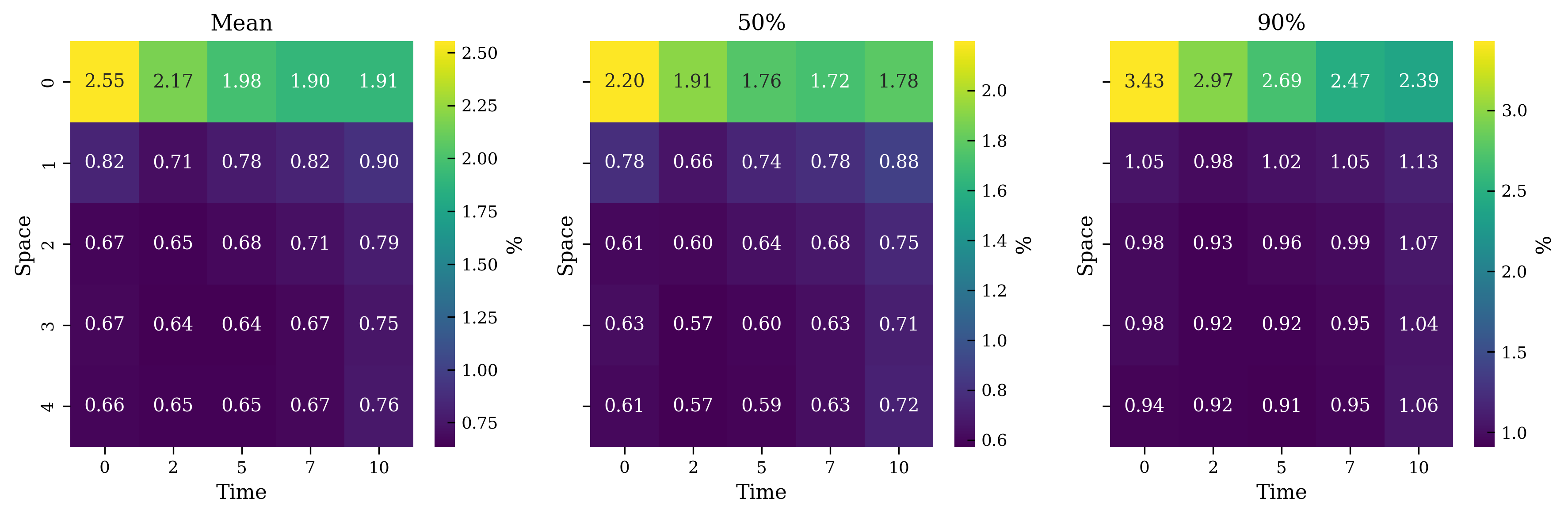}
    \caption{\modifa{Model performance across different spatial locations ($z = \frac{kH}{4} \quad k\in\{0,1,2,3,4\}$ and $t=\frac{k\modifc{t_f}}{10} \quad  k\in \{0,2,5,7,10\}$), using \textbf{MAPE} (in \%) metric and using only 2 thermal images as input to the encoder. The table summarizes the mean and 90th percentile errors of the model, as defined in Eq.~\ref{MAPE_def_eq}, across all validation datasets.}
    }\label{results_APE_chp3}
\end{figure}

\begin{figure}[!htbp]
    \centering
    \includegraphics[width=0.85\textwidth]{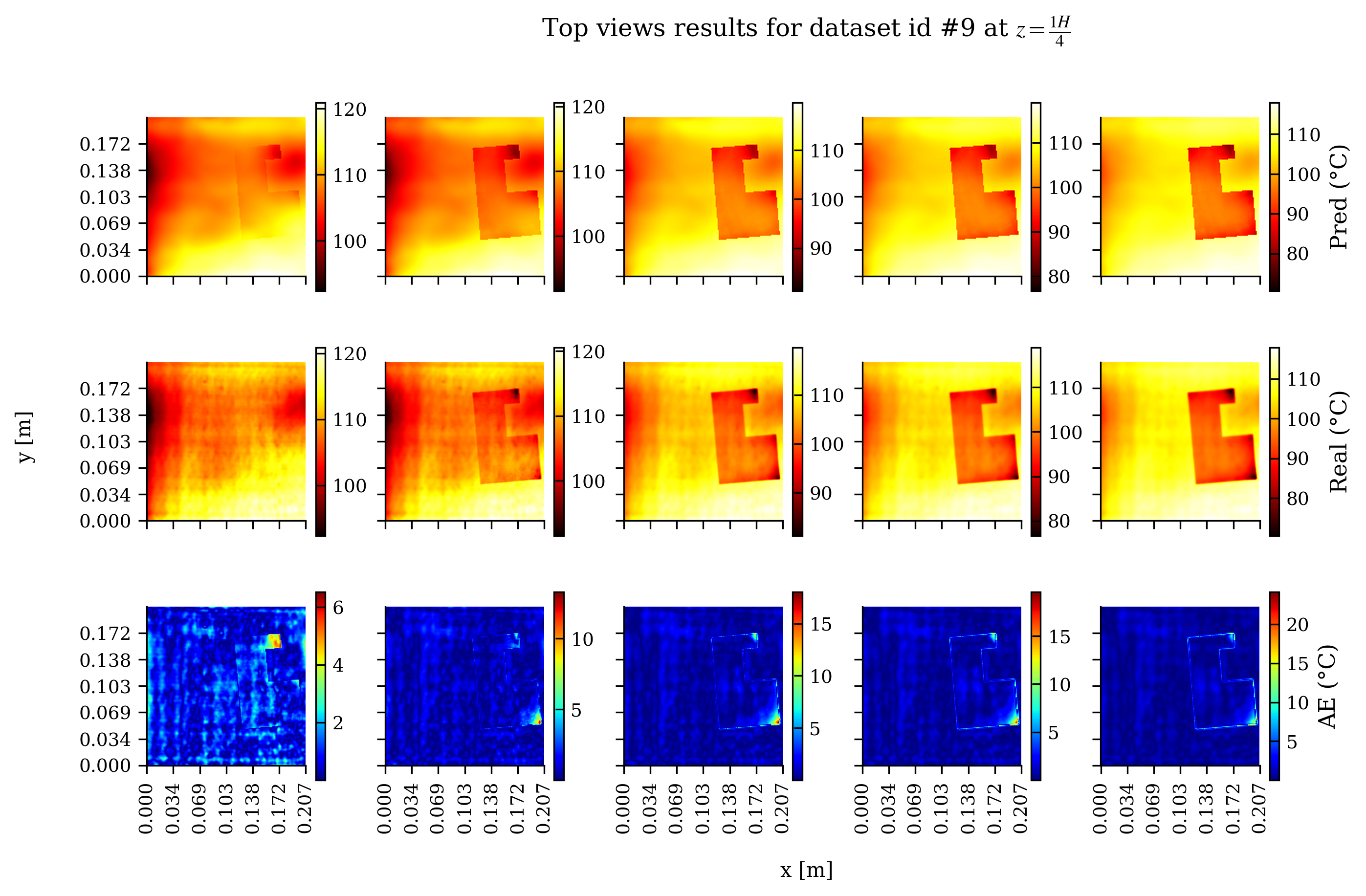}
    \caption{Exemple of model results \modifa{of top view surface temperature} for validation dataset at $z = \frac{H}{4}$ and $t=\frac{k\modifc{t_f}}{10} \quad k\in \{0,2,5,7,10\}$ (columns), the first line represents the model prediction, the second presents the simulation results (ground truth) and the last one represents the Absolute Error between prediction and simulation.
    }\label{results_exp5_z1}
\end{figure}

\begin{figure}[!htbp]
    \centering
    \includegraphics[width=0.85\textwidth]{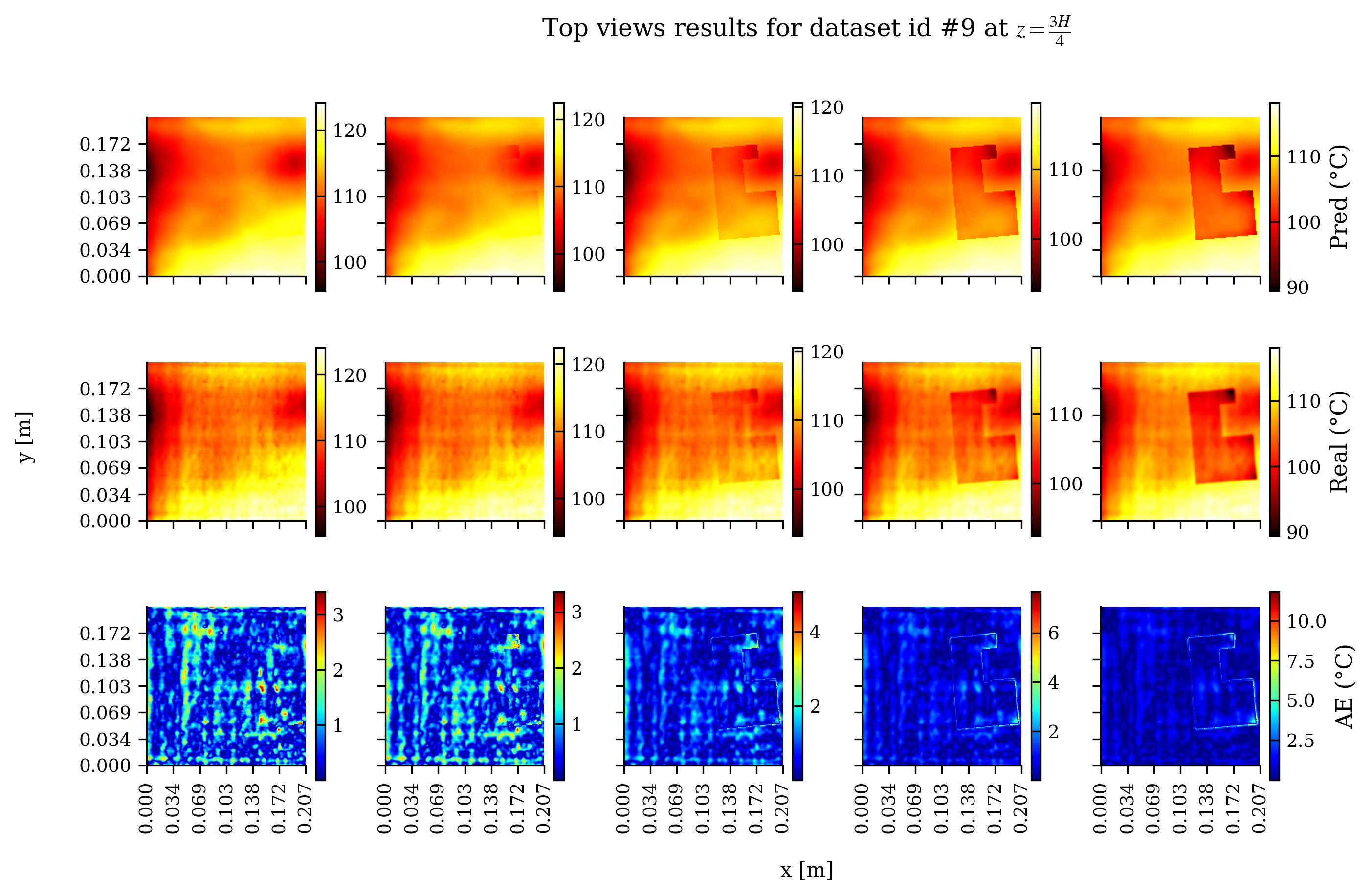}
    \caption{Exemple of model results \modifa{of top view surface temperature} for validation dataset at $z = \frac{3H}{4}$ and $t=\frac{k\modifc{t_f}}{10} \quad k\in \{0,2,5,7,10\}$ (columns), the first line represents the model prediction, the second presents the simulation results (ground truth) and the last one represents the Absolute Error between prediction and simulation.
    }\label{results_exp5_z3}
\end{figure}

The results are given in Fig.~\ref{results_AE_chp3} and~\ref{results_APE_chp3}. Based on these results, it can be seen that the model generalizes well for the validation data and was also able to use the physics to give accurate predictions at $z = \frac{kH}{4} \quad k\in\{1,3\}$, which were not provided to the model during its training. The 90\% percentile could be seen as a confidence factor of the model error; for both metrics, the model error is small. However, the model prediction is better for $z>=\tfrac{H}{4}$, while its prediction is less accurate at $z=0$.
This could be explained by the fact that the temperature difference at the contact surface is high (in the contact region the temperature is low and in the convection region the temperature is high) and also from $z=0$ to $z=\frac{H}{4}$ the temperature difference is very high, which perturbs the model prediction and training. Changing the training strategy of the model, especially by tuning the physics loss parameter $\lambda$ and selecting random collocation points in regions where the residual is big~\citep{HANNA2022115100}, should help improve these results. Some examples of the model prediction versus the simulation results are given in Figs.~\ref{results_exp5_z1} and~\ref{results_exp5_z3} at $z=\frac{H}{4}$ and $z=\frac{3H}{4}$.

\begin{figure}[!htbp]
    \centering
    \includegraphics[width=0.85\textwidth]{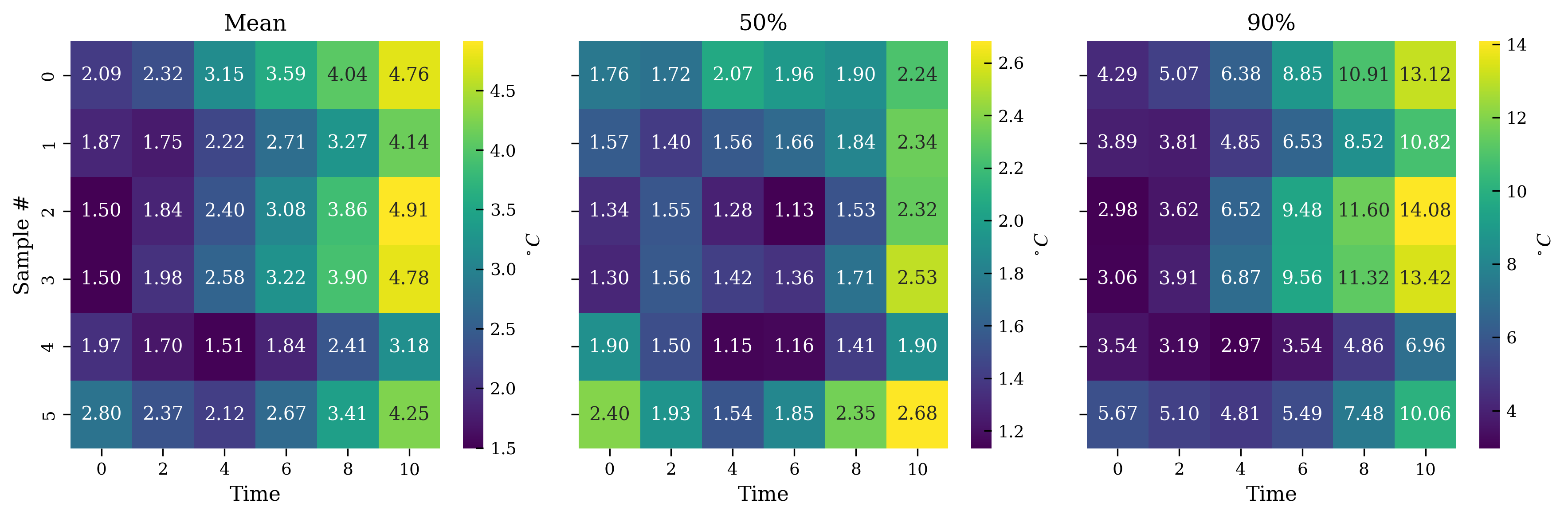}
    \caption{Model performance for \modifa{PA66GF} sheet experiments, across different instances $t=\frac{k\modifc{t_f}}{10} \quad k\in \{0,2,4,6,8,10\}$ and $z = H$, using \textbf{AE} metric and using 4 thermal images as input to the encoder. The figure summarizes the mean, median and 90th percentile of \textbf{AE} of the model as given in Eq.~\ref{AE_def_eq_exp}.
    }\label{AE_results_PA}
\end{figure}

\begin{figure}[!htbp]
    \centering
    \includegraphics[width=0.85\textwidth]{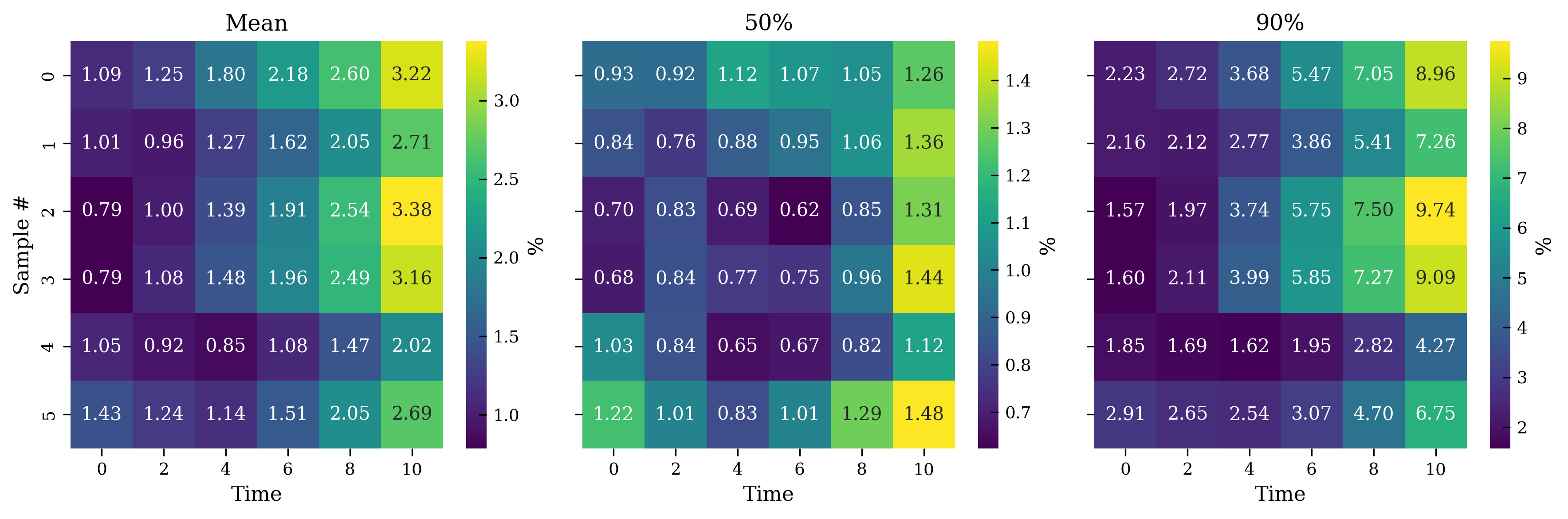}
    \caption{Model performance for \modifa{PA66GF} sheet experiments, across different instances $t=\frac{k\modifc{t_f}}{10} \quad k\in \{0,2,4,6,8,10\}$ and $z = H$, using \textbf{APE} metric and using 4 thermal images as input to the encoder. The figure summarizes the mean, median and 90th percentile of \textbf{APE} of the model as given in Eq.~\ref{APE_def_eq_exp}.
    }\label{APE_results_PA}
\end{figure}

\begin{figure}[!htbp]
    \centering
    \includegraphics[width=0.85\textwidth]{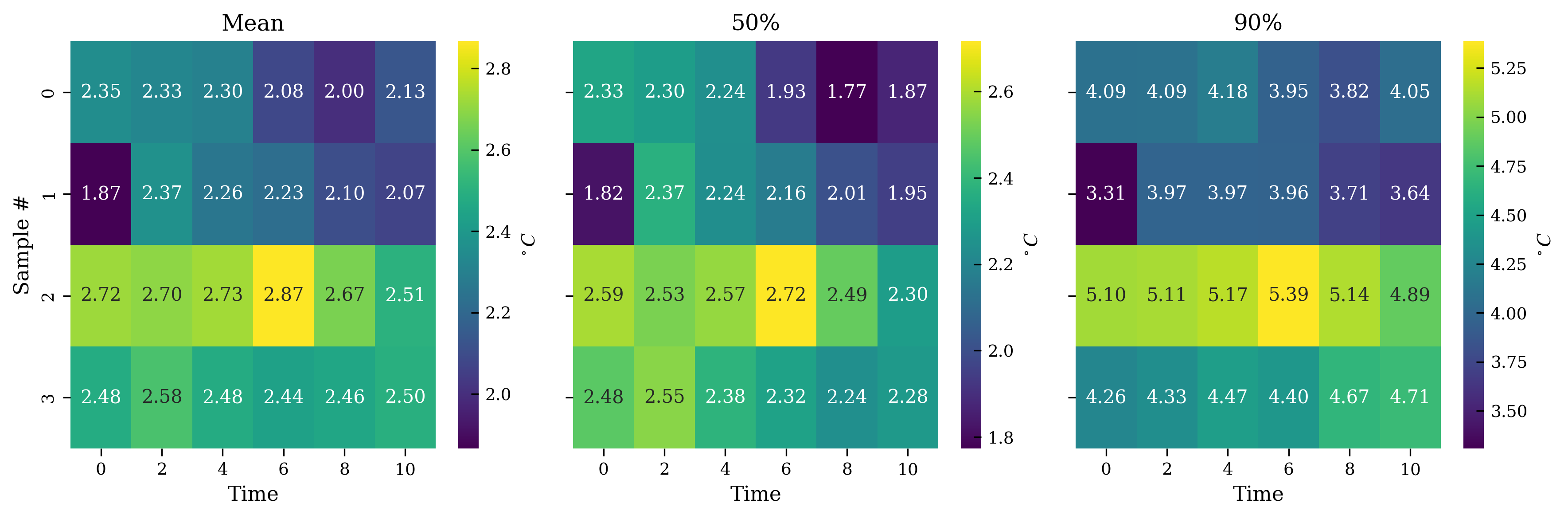}
    \caption{Model performance for \modifa{PPGF} sheet experiments, across different instances $t=\frac{k\modifc{t_f}}{10} \quad k\in \{0,2,4,6,8,10\}$ and $z = H$, using \textbf{AE} metric and using 4 thermal images as input to the encoder. The figure summarizes the mean, median and 90th percentile  of \textbf{AE} of the model as given in Eq.~\ref{AE_def_eq_exp}.
    }\label{AE_results_PP}
\end{figure}

\begin{figure}[!htbp]
    \centering
    \includegraphics[width=0.85\textwidth]{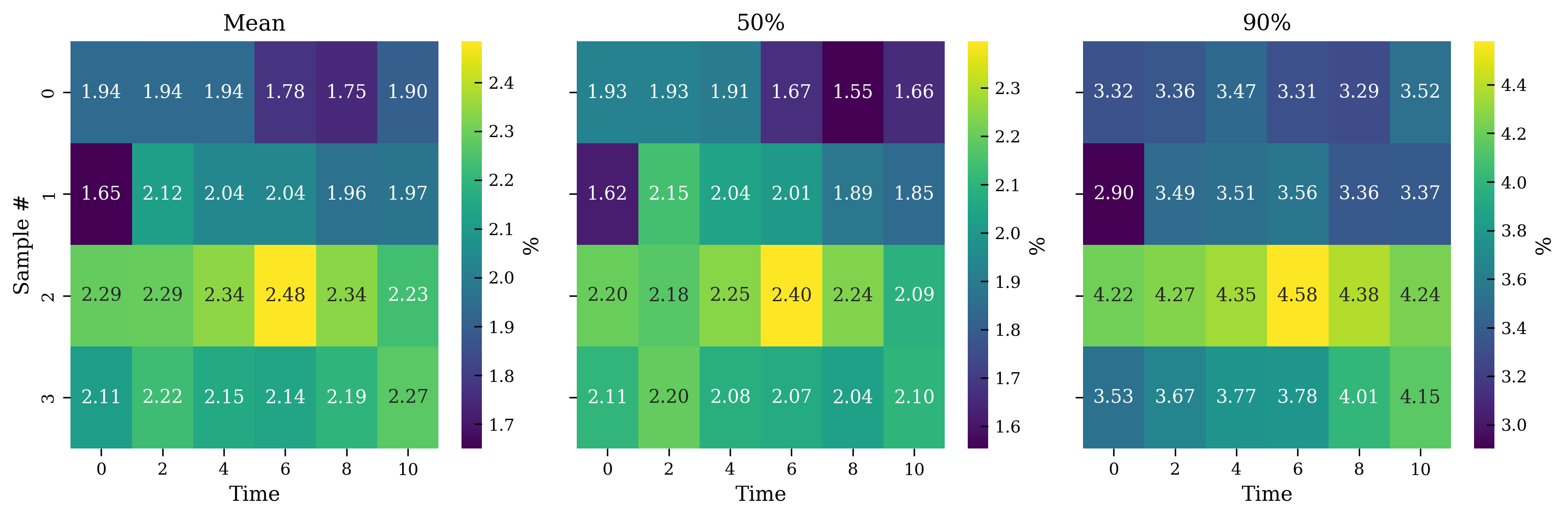}
    \caption{Model performance for \modifa{PPGF} sheet experiments, across different instances $t=\frac{k\modifc{t_f}}{10} \quad k\in \{0,2,4,6,8,10\}$ and $z = H$, using \textbf{APE} metric and using 4 thermal images as input to the encoder. The figure summarizes the mean, median and 90th percentile of \textbf{APE} of the model as given in Eq.~\ref{APE_def_eq_exp}.
    }\label{APE_results_PP}
\end{figure}

\begin{figure}[!htbp]
    \centering
    \includegraphics[width=0.85\textwidth]{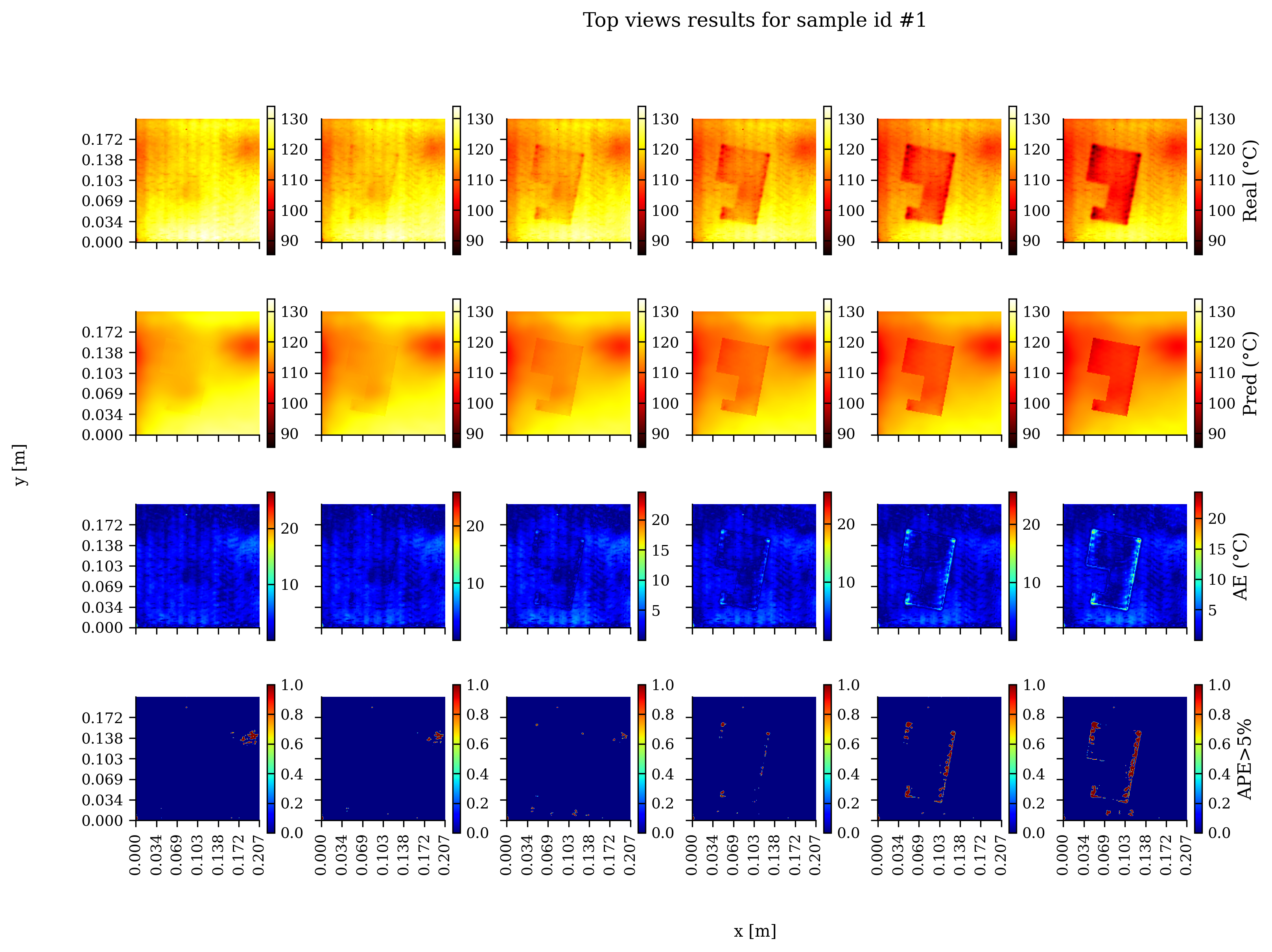}
    \caption{Example of model results \modifa{of top view surface temperature}  for \modifa{PPGF sheet} sample at $z = H$ and $t=\frac{k\modifc{t_f}}{10} \quad k\in \{0,2,4,6,8,10\}$ (columns). The first line represents the thermal images from experiment (ground truth), the second represents the model prediction, the third represents the Absolute Percentage Error (0-1) and the last line represents $\textbf{APE} > 0.12$ (binary values 0 or 1).
    }\label{Example_results_PP_smpl4}
\end{figure}

\paragraph{Validation on real data:} The model is validated on experimental data for all experiments using the same metrics, $\textbf{MAPE}$ and $\textbf{MAE}$ only at $z=H$ (versus data from thermal images) at 6 instances $t_i = \frac{k\modifc{t_f}}{10} \quad k\in\{0,2,4,6,8,10\}$. For each dataset, the $\textbf{MAPE}$, the median and 90\% percentile of $\textbf{APE}$ with respect to surface $\mathbf{x} = (x,y)$ are calculated as given in Eq.~\ref{APE_def_eq_exp} and Eq.~\ref{AE_def_eq_exp}.

\begin{equation}\label{APE_def_eq_exp}
    \begin{aligned}
        \mathrm{APE}(\mathbf{x};~t_i,z = H) &= \frac{|\hat{T}(\mathbf{x},z = H,t_i)-T(\mathbf{x},z = H,t_i)|}{|T(\mathbf{x},z = H,t_i)|}\times 100,\\
        \mathrm{MAPE}_{i,j} &= \frac{1}{|\mathcal{X}|}\sum_{\mathbf{x}\in \mathcal{X}} \mathrm{APE}_{i,j}(\mathbf{x}), \quad d_j \in \mathcal{D}_{\mathrm{Exp}}, \quad j = 1, \dots, N_{\mathrm{val}}\\[1mm] \\
        Q_q(\mathrm{APE}_{j}(t_i,z = H)) &= \inf \Big\{ \eta : \frac{1}{|\mathcal{X}|}\sum_{\mathbf{x}\in \mathcal{X}}\mathbb{I}(\mathrm{APE}_{i,j} \le \eta) \ge q;\quad \mathbf{x}\in \mathcal{X} \Big\}, \; q\in\{0.5,0.9\},\\
    \end{aligned}
\end{equation}

\begin{equation}\label{AE_def_eq_exp}
        \begin{aligned}
            \mathrm{AE}(\mathbf{x};~t_i,z = H) &= |\hat{T}(\mathbf{x},z = H,t_i) - T(\mathbf{x},z = H,t_i)|, \\
            \mathrm{MAE}_{i,j} &= \frac{1}{|\mathcal{X}|}\sum_{\mathbf{x}\in \mathcal{X}} \mathrm{AE}_{i,j}(\mathbf{x}), \quad d_j \in \mathcal{D}_{\mathrm{Exp}}, \quad j = 1, \dots, N_{\mathrm{val}}\\[1mm] \\
            Q_q(\mathrm{AE}_j(t_i,z = H)) &= \inf \Big\{ \eta : \frac{1}{|\mathcal{X}|}\sum_{\mathbf{x}\in \mathcal{X}}\mathbb{I}(\mathrm{AE}_{i,j} \le \eta) \ge q;\quad \mathbf{x}\in \mathcal{X} \Big\}, \; q\in\{0.5,0.9\},\\
        \end{aligned}
\end{equation}
The results using the \modifa{PA66GF} sheet experiments are represented in \modifa{Figs.~\ref{AE_results_PA} and~\ref{APE_results_PA}}, while results for the \modifa{PPGF} sheet are represented in \modifa{Figs.~\ref{AE_results_PP} and~\ref{APE_results_PP}}. An example of the results is given in Fig.~\ref{Example_results_PP_smpl4}.

The model prediction can give some insight about the sheet temperature during the stamping phase and thus could be used to predict regions where temperature is not suitable for the stamping process. In general, the model accuracy based on the real experiment data is good, especially outside the surface of contact for \modifa{PA66GF} sheet. The model was able to accurately encode the ICs pattern (\modifa{thermal images}) into its prediction. While it can be seen that the model was able to map the quality of contact and the mold position into its prediction, it was not able to accurately predict the temperature near the mold edges. This could be due to the inverse problem used to generate the QC, because the simulation results showed similar patterns compared to the experiment results and also due to the physics loss convergence, which needs further investigation to improve its convergence. Furthermore, the results for \modifa{PPGF sheet} are very good compared to \modifa{PA66GF sheet} results. One reason of this difference could be due to the temperature range: while \modifa{PPGF sheet} has a range of $70{-}140\,^{\circ}\mathrm{C}$, \modifa{PA66GF sheet} has a higher range of $100{-}230\,^{\circ}\mathrm{C}$, which makes the prediction more challenging.

\clearpage

\section{Summary and Conclusion}\label{Conc_part}
In this work, the PINN-SE model was tested on the \modifa{heating and early cooling step} of the thermo-stamping process of thermoplastic composite materials, validating the objectives set in this study. The extension of PINN-SE to sequences of 2D thermal images was validated using a combination of CNNs, which encode 2D images into vectors, followed by DeepSet to capture the temporal information within the sequence. The results show that the model can accurately encode \modifa{2D thermal image} sequences alongside other information to predict temperatures in space and time. The extension to multi-encoder inputs was also demonstrated, incorporating different data sources \modifa{such as 2D images, 2D image sequences}, mold geometry and position and other process parameters.

The application of PINN-SE to real data from the thermo-stamping process was demonstrated using an experimental setup that mimics the transfer of the sheet from the oven to the stamping phase, where a thermal camera captures \modifa{the top surface temperature} during natural convection. The model was trained solely on synthetic data generated using FEM simulations, where parameters (ICs/BCs) are sampled from experimental results with the help of a CNN-based VAE model. Results are promising for both synthetic and experimental data, however accuracy dropped for experimental data, especially for \modifa{PA66GF} sheets, likely due to the inverse identification of thermal resistance and the assumption of isothermal TCR without accounting for pressure and gap between sheet and mold. Addressing these limitations by incorporating such parameters in both simulations and the PINN-SE would likely enhance its prediction. Additionally, the QC was used as input to the encoder, which is not easily obtainable in real cases a sequence to image model mapping thermocouple outputs to the QC image could address this limitation.

Despite being trained on a limited dataset with high variability, the model generalized well. However, extending the dataset to more thicknesses and polymers will be necessary for general applicability. Finally, the convergence of the physical loss $\mathcal{L}_{\text{PDE}}$ (Eq.~\ref{loss_TH_4D}) remains computationally expensive, using physics residual-based collocation point sampling~\citep{HANNA2022115100} could reduce computational cost and accelerate convergence.

\appendix

\section{VAE model architecture}\label{app_ch3_3}
\begin{table}[htbp]
    \centering
    \caption{CNN-VAE Architecture for ICs Configuration}
    \label{tab:vae_ics_simple}
    \begin{tabular}{|c|c|c|}
    \hline
    \textbf{Stage} & \textbf{Layer} & \textbf{Activation} \\
    \hline
    \multicolumn{3}{|c|}{\textbf{Encoder}} \\
    \hline
    Input & Input $(161,161)$ & -- \\
    Conv2D + MaxPool & 32 filters, 3$\times$3 same padding, pool (2,2) & swish \\
    Conv2D + MaxPool & 64 filters, 3$\times$3 same padding, pool (2,2) & swish \\
    Conv2D + MaxPool & 128 filters, 3$\times$3 same padding, pool (2,2) & swish \\
    Flatten & -- & -- \\
    Dense $\times 4$ & 32 nodes each & swish \\
    Dense & 8 nodes ($z_{\mu}$) & linear \\
    Dense & 8 nodes ($z_{\sigma}$) & linear \\
    Sampling & Sampling function $z$ & -- \\
    \hline
    \multicolumn{3}{|c|}{\textbf{Decoder}} \\
    \hline
    Dense + Reshape & 20$\times$20$\times$128 & swish \\
    UpSampling + Conv2D & 128 filters & swish \\
    UpSampling + Conv2D & 64 filters & swish \\
    UpSampling + Conv2D & 32 filters & swish \\
    Conv2D (final) & 1 filter & sigmoid \\
    Resize & Resize (161,161) & -- \\
    \hline
\end{tabular}
\end{table}

\begin{table}[htbp]
    \centering
    \caption{CNN-VAE Architecture for QC Configuration}
    \label{tab:vae_qt_simple}
    \begin{tabular}{|c|c|c|}
    \hline
    \textbf{Stage} & \textbf{Layer} & \textbf{Activation} \\
    \hline
    \multicolumn{3}{|c|}{\textbf{Encoder}} \\
    \hline
    Input & Input $(91,59)$ & -- \\
    Conv2D + MaxPool & 32 filters, 3$\times$3 same padding, pool (2,2) & swish \\
    Conv2D + MaxPool & 64 filters, 3$\times$3 same padding, pool (2,2) & swish \\
    Conv2D + MaxPool & 128 filters, 3$\times$3 same padding, pool (2,2) & swish \\
    Flatten & -- & -- \\
    Dense $\times 4$ & 32 nodes each & swish \\
    Dense & 8 nodes ($z_{\mu}$) & linear \\
    Dense & 8 nodes ($z_{\sigma}$) & linear \\
    Sampling & Sampling $z$ & -- \\
    \hline
    \multicolumn{3}{|c|}{\textbf{Decoder}} \\
    \hline
    Dense + Reshape & 12$\times$8$\times$128 & swish \\
    UpSampling + Conv2D & 128 filters & swish \\
    UpSampling + Conv2D & 64 filters & swish \\
    UpSampling + Conv2D & 32 filters & swish \\
    Conv2D (final) & 1 filter & sigmoid \\
    Resize & Resize (91,59) & -- \\
    \hline
\end{tabular}
\end{table}

\clearpage

\section{PINN-SE model architecture}\label{app_ch3_4}
\paragraph{Encoder} The encoder architecture for the sequence of thermal images is given in Tab.~\ref{tab:vae_ics_encoder_PINNSE} for the CNN part and in Tab.~\ref{tab:dps_model} for the Deep Sets. For QC the CNN architecture \modifa{is given in Tab.~\ref{tab:vae_qt_encoder_PINNSE}}. For thickness and mold initial temperature, the used MLP architecture is given in Tab.~\ref{tab:ht_model}, while the polygon coordinates are encoded using an other MLP as described in Tab.~\ref{tab:pts_model}. These feature vectors are then concatenated and fed to an MLP with the architecture given in Tab.~\ref{tab:pre_pinn_fts}.

\begin{table}[htbp]
    \centering
    \caption{CNN Encoder for ICs Configuration (PINN-SE).}
    \label{tab:vae_ics_encoder_PINNSE}
    \begin{tabular}{|c|c|c|}
    \hline
    \textbf{Stage} & \textbf{Layer} & \textbf{Activation} \\
    \hline
    Input & Input $(161,161)$ & -- \\
    Conv2D + MaxPool & 32 filters, 3$\times$3 same padding, pool (2,2) & swish \\
    Conv2D + MaxPool & 64 filters, 3$\times$3 same padding, pool (2,2) & swish \\
    Conv2D + MaxPool & 128 filters, 3$\times$3 same padding, pool (2,2) & swish \\
    Flatten & -- & -- \\
    Dense $\times 4$ & 32 nodes each & swish \\
    Dense & 8 nodes ($z_{\mu}$) & -- \\
    \hline
    \end{tabular}
\end{table}
    
\begin{table}[htbp]
    \centering
    \caption{Deep Set Architecture.}
    \label{tab:dps_model}
    \begin{tabular}{|c|c|c|}
    \hline
    \textbf{Stage} & \textbf{Layer} & \textbf{Activation} \\
    \hline
    Input & Input (sequence size, 9 features) & -- \\
    Dense & 64 nodes with L1 regularization layer & swish \\
    Dense & 64 nodes with L1 regularization layer & swish \\
    Dense & 64 nodes with L1 regularization layer & swish \\
    Dense (output) & 32 nodes with L1 regularization layer & -- \\
    Aggregation & Max pooling across time dimension & -- \\
    \hline
\end{tabular}
\end{table}

\begin{table}[htbp]
    \centering
    \caption{CNN Encoder for QC Configuration (PINN-SE).}
    \label{tab:vae_qt_encoder_PINNSE}
    \begin{tabular}{|c|c|c|}
    \hline
    \textbf{Stage} & \textbf{Layer} & \textbf{Activation} \\
    \hline
    Input & Input $(91,59)$ & -- \\
    Conv2D + MaxPool & 32 filters, 3$\times$3 same padding, pool (2,2) & swish \\
    Conv2D + MaxPool & 64 filters, 3$\times$3 same padding, pool (2,2) & swish \\
    Conv2D + MaxPool & 128 filters, 3$\times$3 same padding, pool (2,2) & swish \\
    Flatten & -- & -- \\
    Dense $\times 4$ & 64 nodes each & swish \\
    Dense & 32 nodes ($z_{\mu}$) & -- \\
    \hline
\end{tabular}
\end{table}

\begin{table}[htbp]
    \centering
    \caption{Thickness and initial mold temperature MLP Architecture.}
    \label{tab:ht_model}
    \begin{tabular}{|c|c|c|}
    \hline
    \textbf{Stage} & \textbf{Layer} & \textbf{Activation} \\
    \hline
    Input & Input (2 features) & -- \\
    Dense & 64 nodes with L1 regularization layer & swish \\
    Dense & 64 nodes with L1 regularization layer & swish \\
    Dense (output) & 32 nodes with L1 regularization layer & -- \\
    \hline
    \end{tabular}
    \end{table}

\begin{table}[htbp]
    \centering
    \caption{Polygon coordinates Model Architecture.}
    \label{tab:pts_model}
    \begin{tabular}{|c|c|c|}
    \hline
    \textbf{Stage} & \textbf{Layer} & \textbf{Activation} \\
    \hline
    Input & Input (16 features) & -- \\
    Dense & 64 nodes with L1 regularization layer & swish \\
    Dense & 64 nodes with L1 regularization layer & swish \\
    Dense (output) & 32 nodes with L1 regularization layer & -- \\
    \hline
\end{tabular}
\end{table}

\begin{table}[htbp]
    \centering
    \caption{Post feature encoding MLP Architecture.}
    \label{tab:pre_pinn_fts}
    \begin{tabular}{|c|c|c|}
    \hline
    \textbf{Stage} & \textbf{Layer} & \textbf{Activation} \\
    \hline
    Input & Input (104 features) & -- \\
    Dense & 128 nodes with L1 regularization layer & swish \\
    Dense & 128 nodes with L1 regularization layer & swish \\
    Dense (output) & 64 nodes with L1 regularization layer & linear \\
    \hline
\end{tabular}
\end{table}

\paragraph{Space-time Encoder} The Random Fourier Features layer architecture used to encode the $(x,y)$ spatial data, along with the MLP used to encode the concatenation of the Fourier Features output and $z,~t$ is given in Tab.~\ref{tab:pre_pinn_txyz_full}.

\begin{table}[htbp]
    \centering
    \caption{Space-time Encoder Architecture with Fourier Features layer.}
    \label{tab:pre_pinn_txyz_full}
    \begin{tabular}{|c|c|c|}
    \hline
    \textbf{Stage} & \textbf{Layer} & \textbf{Activation} \\
    \hline
    
    Fourier Features & Random Fourier Features, 30 nodes (scale = 0.1) & -- \\
    Concatenate & -- & -- \\
    Input & Input (32 features) & -- \\
    Dense & 100 nodes with L1 regularization layer & swish \\
    Dense & 100 nodes with L1 regularization layer & swish \\
    Dense (output) & 32 nodes with L1 regularization layer & -- \\
    \hline
\end{tabular}
\end{table}

\paragraph{PINN} Two identical MLPs are used to represent the PINN architecture, as given in Tab.~\ref{tab:pinn_mlp}.

\begin{table}[htbp]
    \centering
    \caption{The PINN Architecture.}
    \label{tab:pinn_mlp}
    \begin{tabular}{|c|c|c|}
    \hline
    \textbf{Stage} & \textbf{Layer} & \textbf{Activation} \\
    \hline
    Input & Input (86 features) & -- \\
    Dense & 100 nodes with L1 regularization & swish \\
    Dense & 100 nodes with L1 regularization & swish \\
    Dense & 100 nodes with L1 regularization & swish \\
    Dense & 100 nodes with L1 regularization & swish \\
    Dense & 100 nodes with L1 regularization & swish \\
    Dense (output) & 1 nodes with L1 regularization layer & -- \\
    \hline
\end{tabular}
\end{table}
\clearpage

\section*{Acknowledgements}
For the purpose of Open Access, a CC-BY public copyright licence has been applied by the authors to the present document and will be applied to all subsequent versions up to the Author Accepted Manuscript arising from this submission.
\section*{Funding}
The authors would like acknowldge the funding and support of the PERFORM Program of IRT Jules Verne.

\bibliographystyle{elsarticle-num} 
\bibliography{elsarticle-template-num}

\end{document}